\documentclass{aa}  
\usepackage{natbib}
\bibpunct{(}{)}{;}{a}{}{,} 
\newcommand\ttt{\text}
\newcommand\mrm{\mathrm}

\usepackage{graphicx}
\usepackage[varg]{txfonts}

\usepackage[colorlinks=true, allcolors=blue]{hyperref}
\newcommand{\mLE}[1]{{#1}}
\newcommand{\rLE}[1]{}

\begin{document} 

 \title{Computation of the lateral shift due to atmospheric refraction}
    \author{H. Labriji  \and 
          O. Herscovici-Schiller  \and 
          F. Cassaing
          }

   \institute{DTIS, ONERA, Université Paris Saclay \\ F-91123 Palaiseau - France \\
   \email{hanae.labriji@onera.fr}    }


  \abstract
   {Atmospheric refraction modifies the apparent position of objects in the sky. As a complement to the well-known angular offset, we computed the lateral translation that is to be considered for short-range applications, such as wavefront sensing and meteor trajectories.
   }
   {We aim to calculate the lateral shift at each altitude and study its variation according to meteorological conditions and the location of the observation site. We also pay special attention to the chromatism of this lateral shift. Moreover, we assess the relevance of the expressions present in the literature\mLE{, which} have been established neglecting Earth’s curvature\rLE{all the expressions present in the literature are neglecting Earth's curvature}.  
    }
   {We extracted the variation equations of refraction from the geometric tracing of a light ray path. A numerical method and a dry atmosphere model allowed us to numerically integrate the system of coupled equations. In addition to this, based on Taylor expansions, we established three analytic approximations of the lateral shift, one of which is the one already known in the literature. We compared the three approximations to the numerical solution. All these estimators are included in a \textsc{Python 3.2} package, which is available online.}
  {Using the numerical integration estimator, we calculated the lateral shift values for any zenith angle including low elevations. The shift is typically around $3\,\ttt{m}$ at a zenith angle of $45^\circ$, $10\,\ttt{m}$ at $65^\circ$, and even $300\,\ttt{m}$ at $85^\circ$. Next, the study of the variability of the lateral shift as a function of wavelength shows differences of up to 2\% between the visible and near infrared. Furthermore, we show that the flat Earth approximation of the lateral shift corresponds to its first-order Taylor expansion. The analysis of the errors of each approximation shows the ranges of validity of the three estimators as a function of the zenith angle. The `flat Earth' estimator achieves a relative error of less than 1\% up to $55^\circ,$ while the new extended second-order estimators improves this result up to $75^\circ$.}
   {The flat Earth estimator is sufficient for applications where the zenith angle is below $55^\circ$ (most high-resolution applications) but a refined estimator is necessary to estimate meteor trajectories at low elevations.}


   \maketitle

\section{Introduction}

\begin{figure}[!ht]
    \centering
    \includegraphics[width=0.45\textwidth,trim={18cm 1cm 20cm 1cm},clip]{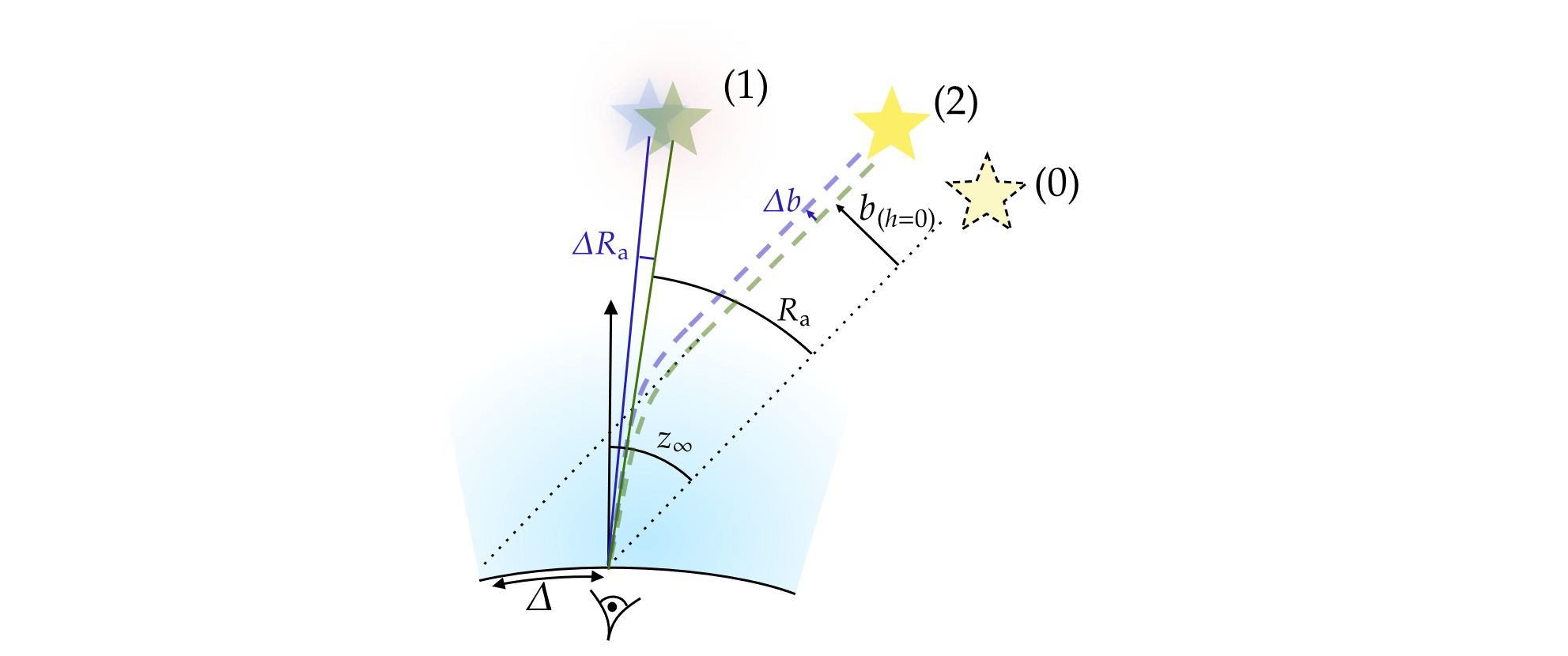}
    \caption{Bending of light rays due to atmospheric refraction.}
    \label{fig:refraction_intro}
\end{figure}

Due to the decrease in the air refractive index with altitude, light rays entering the atmosphere or emitted from Earth are bent, respectively, towards the ground or towards the horizon. This phenomenon, illustrated in Fig.~\ref{fig:refraction_intro}, has spatial and temporal consequences. First, the ray is bent towards the ground, making the actual ray differ from the expected ray. The deviation between the two trajectories is characterised by a shift $b$ outside the atmosphere, the computation of which is the main goal of this paper, and a difference in the apparent position of the object at the observer level. The deviation between the true {geometric} direction (tag (0)) and the apparent {optical} direction (tag (1)) is the well-known angle of refraction $R_a$ \citep{mahan_astronomical_1962, thomas1996astronomical} and depends on the observation wavelength, which causes a visible stretching of the stars $\Delta R_a,$ also known as the blur effect. Second, the ray distortion changes the  trip time across the atmosphere, which is of interest for satellite ranging systems \citep{gardner_correction_1977, dodson_refraction_1986, abshire_atmospheric_1985, marini_correction_1972}.

The purely angular correction $R_\ttt{a}$ is well adapted to astro\-metry, \mLE{namely} when compensating the position of an object optically located at infinity such as a star. However, there are a number of cases where it is necessary to take into account the shift $b$ of the ray with respect to its vacuum path {(tag (2))} and its chromatism $\Delta b$, as pictured in Fig.~\ref{fig:refraction_intro}.

A first case is the correction of atmospheric turbulence over a wide spectral domain. It is usual in adaptive optics to operate the wavefront sensor in a different band than the imaging sensor \citep{nakajima_zenith-distance_2006}. Stellar interferometers are also impacted by this issue, as they can operate on different bands, while only one of them is dedicated to the fringe tracker \citep{pannetier-longDispCorrection-mnras21}. The chromaticity of the lateral shift (Fig.~\ref{fig:refraction_intro}), sometimes called the `chromatic shear' \citep{nakajima_zenith-distance_2006}, makes the integrated wavefront error vary with wavelength, creating a `chromatic anisoplanatism' \citep{devaney_chromatic_2008,sasiela_strehl_1992} that is not corrected even if an atmospheric dispersion corrector is used to correct dispersion in the focal plane. As it is difficult to correct the impact of refraction, it is mostly considered as part of the residual errors induced by the atmospheric dispersion corrector  \citep{van_denborn_quantification_2020}.
 
The second case is estimating the position of an object near Earth using its apparent position in the sky, as typically used in meteor tracking networks \citep{mccrosky_special_1968, jeanne_calibration_2019, gardiol_cavezzo_2021, colas_fripon_2014, colas_fripon_2020, egal_challenge_2017, vida_high_2021, TOTH2015102}.  Since the object is generally observed at an altitude larger than $20 \, \mrm{km}$, its apparent position is affected at low elevations by atmospheric refraction. The all-sky cameras calibrate the position of the identified objects based on the known positions of the stars \citep{borovicka_new_1995}. Yet, the fact that the object is not at an infinite distance from the observer (unlike the stars) makes this correction unsuitable, and we can see in Fig.~\ref{fig:refraction_intro} that the error is equal to the lateral shift $b$.

The third example of the shift compensation's usefulness is the observation of Earth from space. This time, the projection $\Delta$ of the lateral shift on the ground (Fig.~\ref{fig:refraction_intro}) is used to compensate refraction when estimating the position of the generated data \citep{yan_correction_2016,li_method_2016}. Thus, since atmospheric wavefront distortions are much less critical when observing Earth from space than space form Earth, Earth observation satellites can operate at large zenith angle and have their line of sight significantly shifted on the ground. 

Atmospheric refraction was recorded in the literature very early, and Aristotle mentions the vertical stretching at the horizon in his book Meteorology \citep{aristotle_works_1908}, while the first instrumental proof was made by Johannes Schöner and reported in a monograph written by \cite{regiomontanus_scripta_1544}. The famous Danish astronomer Tycho Brahe was one of the first people to measure the effect of refraction using the apparent position of sun at Earth's summer and winter solstices \citep{mahan_astronomical_1962}. 

Later, the angle $R_\ttt{a}$ has been the subject of several studies; authors such as \cite{radau_recherches_1882}, \cite{saastamoinen_contributions_1972}, and \cite{chambers_astrometry_2005} have approached the refraction angle with a series of approximations, while others integrated the variation equations of the light path numerically \citep{ hohenkerk_computation_1985,seidelmann_explanatory_1992,stone_accurate_1996,wittmann_astronomical_1997,  kristensen_astronomical_1998,auer_astronomical_2000,  nauenberg_atmospheric_2017}. In contrast to the refraction angle, the calculation of the lateral shift $b$ has been identified as a difficult problem by \cite{mccrosky_special_1968}, and there have been a few early attempts to approximate it \citep{schmid_influence_1963-1}. Within the framework of adaptive optics, Wallner has written two papers on this issue \citep{wallner_effects_1976,wallner_minimizing_1977}. In \cite{wallner_effects_1976}, one can find an expression of the lateral shift (as a function of the altitude $h$) whose derivation assumes that the Earth is flat: 
\begin{equation}\label{equ:b_wallner}
    b(h,\lambda) = N_\ttt{S}(\lambda)\frac{\tan{z_{\infty}}}{\cos{z_{\infty}}} \frac{P(h)}{g \rho_\ttt{S}}\,,
\end{equation}
where $z_\infty$ refers to the zenith angle of the non-refracted ray, $P$ is the atmospheric pressure at altitude $h$, $g$ is the standard gravity, $\rho_\ttt{S} = 1.225\,\mrm{kg/m^3}$ is a constant, and $N_\ttt{S}(\lambda)$ represents the variation of the refractive index with wavelength ($\lambda$).

Based on Eq.~(\ref{equ:b_wallner}), some authors have continued his work to evaluate wavefront sensing degradation for extremely large telescopes (ELTs) \citep{owner-petersen_consequences_2004,owner-petersen_effects_2006,jolissaint_modeling_2010}. Several other papers seem to have independently found and used the same expression of the lateral shift \citep{sasiela_strehl_1992,nakajima_zenith-distance_2006}. 

This article is dedicated to the calculation of the lateral shift. We begin Sect.~\ref{sec:section_2} by rigorously writing the differential equations in the spherical Earth model. The remaining part of Sect.~\ref{sec:section_2} is devoted to the study of the estimator (denoted $b^{(\infty)}$) resulting from the numerical integration of the local equations. \mLE{The evolution of the estimator is also investigated as a function of the observation wavelength and the temperature and pressure conditions.} Then, in Sect.~\ref{sec:taylor_expansion}, we describe our use of a Taylor expansion of the variational equations to deduce three other estimators of the lateral shift $b^{(1)}$ (Eq.~(\ref{equ:approx_shift_1})), $b^{(3/2)}$ (Eq.~(\ref{equ:b32})), and $b^{(2)}$ (Eq.~(\ref{equ:b2_approx})). We provide the means to calculate their values knowing only the meteorological conditions of the observation site and the distribution of air in the atmosphere.  In Sect.~\ref{sec:accuracy_estimators}, we study the accuracy of each approximation as a function of the zenith angle.  Finally, in Sect.~\ref{sec:application} we discuss two applications of the lateral shift. In
Sect.~\ref{sec:lateral_shift_OA}, we start by studying the effect of numerical integration on the `chromatic shear' values in wavefront sensing, and then we focus on the effect of the lateral shift for the observation of nearby objects (Sect.~\ref{sec:lateral_shift_nearbyObjects}). Theoretical models for the atmosphere and its moments are described in the appendices. A companion \textsc{Python} program to this article is available\footnote{Labriji, H. 2021, \url{https://github.com/OneraHub/RefractionShift} \label{footnote:github}}.

\section{Derivation of the lateral shift in the spherical atmosphere model}\label{sec:section_2}

\subsection{Differential equations from geometric analysis}
\label{subsec:equation_integration}

\begin{figure}[!ht]
    \centering
    \includegraphics[width=0.49\textwidth,trim={1cm 0cm 1.5cm 0.3cm},clip]{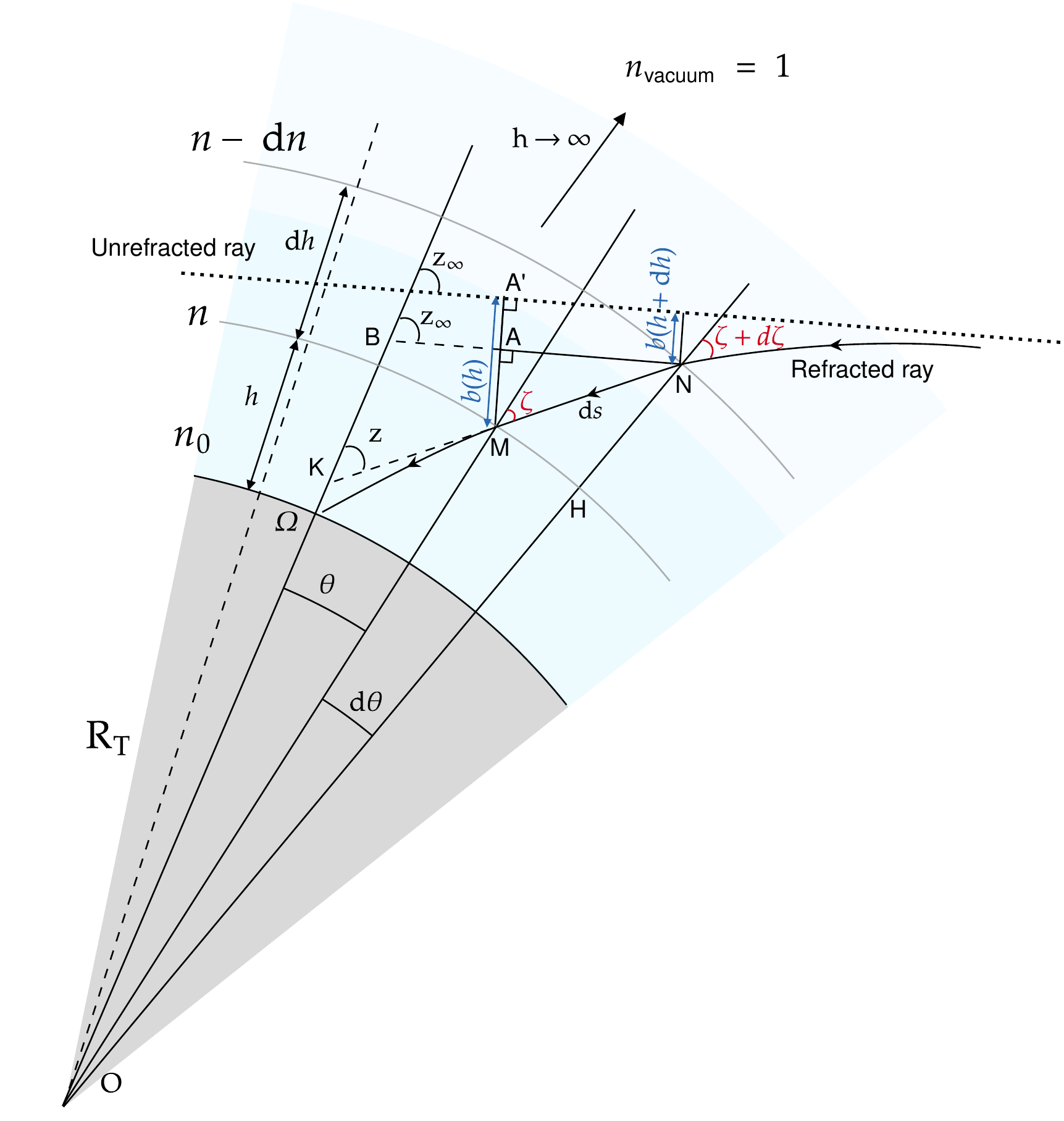}
    \caption{Geometric notations.}
    \label{fig:devLat_notation_ds}
\end{figure}

We consider a ray that reaches the observer at point $\Omega$ in a spherical model of Earth's atmosphere as depicted in Fig.~\ref{fig:devLat_notation_ds}. The light ray passes through point $M$ at the altitude $h$ and point $N$ at $h+\mathrm{d}h$. The hypothetical path in absence of atmosphere is represented by a dotted line. We note \mLE{$R_\ttt{T}$ the Earth's radius, $h$ the altitude of the point $M$, $n$ the local refractive index, $z$ the angle that the ray makes at a point $M$  with the direction of the zenith at the observer, $\zeta$ the local zenith angle at each altitude, and $s$ the length along the optical path.} 

The subscript 0 always refers to a quantity measured at ground level (i.e. at the observer level), while the subscript $\infty$ refers to a quantity at infinity. $z_0$ is the apparent zenith angle at the observer level, and $z_\infty$ is the angle between the non-refracted ray and the zenith at $\Omega$. The classical refraction angle is denoted $R_\ttt{a,}$ and:
\begin{equation}
    R_\ttt{a} = z_\infty - z_0.
\end{equation} 

The variation of the local zenith angle at each altitude $h$ sa\-tisfies the Bouguer invariant and depends on the local refractive index as follows:
\begin{equation}\label{equ:bouguer_invariant}
    R_\ttt{T}\, n_0 \sin{\zeta_0} =(R_\ttt{T}+h) \, n \sin{\zeta}.
\end{equation}
This formula stems from Snell Descartes' law; its proof can be found in Chapter 6 of \cite{kovalevsky_fundamentals_2004}. 

In the following, we derive the equations of variations of the main variables describing the optical path of the light ray: $z$, $\zeta,$ and $h$. Some of those are already known in the usual computation of the angle of refraction $z_\infty -z_0$ \citep{hohenkerk_computation_1985}. The goal here is to express the lateral shift $b(h)$ according to given parameters such as the zenith angle. As already highlighted in many references \citep{kovalevsky_fundamentals_2004}, the main difficulty of this calculation in spherical atmosphere is the continuous change of the zenith direction ($\zeta$) along the ray path. Using the variable $z$, which represents the angle that the ray makes with the zenith at the point of observation, allows us to overcome this difficulty. 

First, differentiating the Bouguer formula, one obtains 
\begin{equation}\label{equ:diff_bouguer}
     \frac{\mathrm{d}h}{R_\ttt{T}+h} \, \tan{(\zeta)}+ \mathrm{d}\zeta = -  \frac{\mathrm{d}n}{n} \, \tan{(\zeta)}.\end{equation}

Then, introducing $\theta$ in Fig.~\ref{fig:devLat_notation_ds}, one has the following in the $OKM$ triangle\rLE{ I added a space between 'OKM' and 'triangle'}:
\begin{equation}\label{equ:z_thet_zeta}
    z = \zeta + \theta.
\end{equation}

Moreover, in the $MHN$ triangle (where $MH$ is a circle arc): 
\begin{equation}\label{equ:equ0}
    (R_\ttt{T} + h) \, {\mathrm{d}\theta} = \tan{(\zeta)} \, \mathrm{d}h,
\end{equation}
which implies, combined with Eq.~(\ref{equ:diff_bouguer}),
\begin{equation}\label{equ:dz_2}
    \mathrm{d}z = - \frac{\mathrm{d}n}{n} \, \tan{(\zeta)}.
\end{equation}

To obtain the values of $z$, $\theta,$ and $\zeta$ along the optical path numerically, it is necessary to integrate the variation equations according to a chosen integration variable. There are several possible integration variables. Knowing $n$ and $\mathrm{d}n/ \mathrm{d}h$, the choice of $h$ as the integration variable is the most intuitive, but it reveals a singularity near the horizon because $\tan\,(\zeta)$ in Eq.~(\ref{equ:dz_2}) is no longer defined at $90^\circ$. \cite{auer_astronomical_2000} recommend using $z$ as the integration variable, but this adds a computational step and makes a singularity appear at the zenith \citep{hohenkerk_computation_1985}. We choose to use $s$ here, as recommended by \cite{van_der_werf_comment_2008}, because it avoids singularities at both ends.

Considering again the $MHN$ triangle, we can connect the infinitesimal variations of $s$ to $h$ using 
\begin{equation}\label{equ:dhds}
    \mrm{d}h = \cos{(\zeta)} \, \mrm{d}s.
\end{equation}
From Eqs.~(\ref{equ:dz_2}) and (\ref{equ:dhds}), we obtain: 
\begin{equation}\label{equ:dz_3}
    \mathrm{d}z = - \frac{\sin{(\zeta)}}{n} \frac{\mathrm{d}n}{\mathrm{d}h} \, \mathrm{d}s,
\end{equation}
and from Eqs.~(\ref{equ:equ0}) and (\ref{equ:dhds}), we also obtain:
\begin{equation}\label{equ:dtheta}
    \mathrm{d}\theta = \frac{\sin{(\zeta)}}{R_\ttt{T} + h} \, \mathrm{d}s.
\end{equation}

We deal next with the lateral shift $b$. Following Fig.~\ref{fig:devLat_notation_ds}, where $A'$ is the orthogonal projection of $M$ on the non-refracted ray, and $A$ is the orthogonal projection of $N$ on $(MA')$, the lateral shift of the ray from its non-refracted path at an altitude $h$ is defined positively as
\begin{equation}
    b(h) = A'M.
\end{equation}
The reference of $b$ can be chosen either at ground level or at infinity, depending on the purpose of the calculation. In what follows, we choose the reference of $b$ at infinity: $b(h\rightarrow\infty) = 0$. Since the reference of $h$ and $s$ are at sea level, the lateral shift decreases according to the altitude and $\mathrm{d}b \leq 0$. Due to the latter, and considering the triangles $AMN$ and $BNK$, one can write 
\begin{equation}\label{equ:equdb}
    \mrm{d}b = b(h+\mrm{d}h) - b(h) = -AM = \sin{\left( z-z_\infty \right)} \, \mrm{d}s  . 
\end{equation}

Since the angle of refraction along the ray $\left( z_\infty - z(h) \right)$ is always positive, the sine term in Eq.~(\ref{equ:equdb}) supports the fact that the lateral shift $b$ decreases along the optical path.

Finally, gathering Eqs.~(\ref{equ:dhds}) to~(\ref{equ:dtheta}), (\ref{equ:z_thet_zeta}), and~(\ref{equ:equdb}), we end up with a system of five first-order coupled non-linear ordinary differential equations:
\begin{subequations}\label{equ:system_refraction_ds_2}
    \begin{align}
        \frac{\mathrm{d}h}{\mathrm{d}s} = & \cos{(\zeta)}, \label{equ:sys1}\\
        \frac{\mathrm{d}z}{\mathrm{d}s } = &- \frac{\sin{(\zeta)}}{n} \frac{\mathrm{d}n}{\mathrm{d}h}, \label{equ:sys2}\\
        \frac{\mathrm{d}\theta}{\mathrm{d}s} = & \frac{\sin{(\zeta)}}{R_\ttt{T} + h},\label{equ:sys3}\\
        \frac{\mathrm{d}\zeta}{\mathrm{d}s} = & -\frac{\sin{(\zeta)}}{n} \frac{\mathrm{d}n}{\mathrm{d}h} - \frac{\sin{(\zeta)}}{R_\ttt{T} + h} , \label{equ:sys4}\\
        \frac{\mathrm{d}b}{\mathrm{d}s} =  & \sin{\left(z - z_{\infty}\right)}. \label{equ:sys5}
    \end{align}
\end{subequations}
Of the five variables, one can choose to omit the variable $\theta$ for efficiency reasons and because it is linked to $z$ and $\zeta$ by Eq.~(\ref{equ:z_thet_zeta}). Yet, we chose to keep it since it is needed in Sect.~\ref{sec:lateral_shift_nearbyObjects}.

\subsection{The refractive index model}\label{sec:dry_atm}
In order to solve the system of Eqs.~(\ref{equ:system_refraction_ds_2}), it is necessary to provide $n$ and $\mathrm{d}n/\mathrm{d}h$ over the entire height of the atmosphere. Since the true values of the refractive index are not easily obtained, it is common to choose a particular model of the atmosphere. The choice of an appropriate theoretical model depending on the environment of the observation is very important, as it can significantly affect the refraction angle. For instance, \cite{lovchy_calculation_2021} found differences of up to $20\%$ on the angle of refraction at $z_0=90^\circ$ between eight different models. Even when the right simplified model is chosen, there are always substantial deviations between the modelled temperature profile and reality. \cite{nauenberg_atmospheric_2017} compared the calculation of refraction using the standard profile and probe balloon measurements, and he found discrepancies of around one arcsecond for small zenith angles and even a few arcminutes from the horizon (Table 2 and 3).

Here, we choose to follow an approximation made by se\-veral authors for the computation of the angle of refraction \citep{hohenkerk_computation_1985, neda_flatness_2002, auer_astronomical_2000}. Their model truncates the US76 standard atmospheric profile that is based on a piecewise linear temperature for nine atmospheric layers \citep{coesa_us_1976}. The choice to simplify it and reduce it to only two layers is motivated by the fact that most of the atmospheric refraction occurs in the troposphere, since the air beyond it is very sparse. This simplified model is also validated by the good match of computed results with the observational data (such as sunrise and sunset times \citep{neda_flatness_2002}).
Earth's atmosphere is assumed to be a perfect gas \mLE{of} low density \mLE{in} hydrostatic equilibrium\rLE{ using in terms of changes the scientific meaning}. It follows the Gladstone Dale relation \citep{gladstone_xiv_1863}: 
\begin{equation}\label{equ:gladstone_dale_1}
    n-1 = \kappa \rho ,
\end{equation}
where $\kappa$ is a parameter that depends on the observation wavelength $\lambda$, and $\rho$ is the local density of air. In order to express the density $\rho$ as a function of altitude, it is sufficient to compute a temperature profile based on a constant temperature gradient in the troposphere:
\begin{equation}\label{equ:gamma_grandient_T}
    \omega = {\mathrm{d}T}/{\mathrm{d}h} = -6.5 \text{ K/km,}
\end{equation}
and a constant temperature beyond the tropopause (i.e. the temperature gradient is nil). Setting only the value of the temperature gradient $\omega$ allows us to adapt the evolution of the temperature according to its value $T_0$ measured at ground level. Thus, our inputs are the temperature and pressure at ground level (respectively $T_0$ and $P_0$). The value of $\omega$ depends on the chosen model as one can see in the tables published in \cite{Anderson1986} for six reference atmospheres.

Since atmospheric refraction is only slightly sensitive to air moisture (as the values of the angle of refraction in dry and moist air are very close \citep{van_der_werf_ray_2003}), the relative humidity is considered as constant and nil in the troposphere and above it. As for the parameter $\kappa$, it is proportional to the refractivity of dry air $A_d$ and depends on the wavelength $\lambda$. The refractivity of air is also very difficult to measure accurately, several different formulae have been proposed by different authors \citep{barrell_refraction_1939,edlen_dispersion_1953, edlen_refractive_1966, owens_optical_1967}. \cite{ciddor_refractive_1996} gave the most recent expression of refractivity in dry and moist air, also regarded as the most accurate formulae. Yet, only very few authors have compared the refraction angles for different refractivities with true experimental data. \cite{skemer_direct_2009} used spectrography to compare atmospheric dispersion in the N band using Mathar’s model \citep{mathar_refractive_2007} with actual on-sky measurements; the model was in good agreement with the data. Recently, \cite{wehbe_-sky_2020} suggested a new spectrography method to measure on-sky atmospheric dispersion, showing experimentally in \cite{wehbe_-sky_2021} that, apart from the model used by \cite{hohenkerk_computation_1985}, all the used models offer residual dispersions lower than $20\,\ttt{milliarcseconds}$ in the $315-665\,\mrm{nm}$ wavelength range. 

Therefore, in order to minimise the errors due to the model, we have chosen Ciddor’s and Mathar’s refractivity models and kept both the temperature and pressure at the observer as input parameters, to be as close as possible to reality. In Appendix~\ref{app:dry_atmosphere} , we provide the details of the calculation of $n$ and $\mrm{d}n/\mrm{d}h$ at each altitude.

\subsection{Runge-Kutta estimator of the lateral shift: \texorpdfstring{$b^{(\infty)}$}{Lg}}\label{sec:numerical_method}

Now that we have chosen a model for the refractive index of Earth's atmosphere, we next present the numerical arguments of solving the system of coupled ordinary differential Eqs.~(\ref{equ:system_refraction_ds_2}). This resolution provides us with a first estimator that we refer to as $b^{(\infty)}(h)$. The infinity exponent is related to the existence of several finite-order approximations developed in Sect.~\ref{sec:taylor_expansion}. In the following, we denote the values of the lateral shift obtained by evaluating
the estimator $b^{(\infty)}$ at $h = h_0$ by $b_0$.

To solve Eqs.~(\ref{equ:system_refraction_ds_2}), we first notice that the ${\mrm{d}b}/{\mrm{d}s}$ variation equation also requires the value of $z_\infty$. Therefore, it is necessary to proceed in two separate steps: \mLE{we first solve a first equation system (Eqs.~(\ref{equ:sys1}) to~(\ref{equ:sys4})) to calculate the refraction angle $R_\ttt{a}$; then, we solve the complete system (Eqs.~(\ref{equ:sys1}) to~(\ref{equ:sys5})) to derive the lateral shift $b$.}

Both computations are made using a fourth-order Runge-Kutta method. The first one is solved in the reverse direction of the optical propagation, while the second one is done in the direction of the optical propagation. We choose a constant integration step equal to $100\,\ttt{m}$. This value is justified in Sect.~\ref{subsec:integration_step}. 

For the first equation system (Eqs.~(\ref{equ:sys1}) to~(\ref{equ:sys4})), we \mLE{placed} ourselves in the point of view of the observer and \mLE{considered} the following initial state:
\begin{equation}
   \begin{bmatrix}h \\ z \\ \theta \\ \zeta \end{bmatrix}_{h = h_0} = \begin{bmatrix}h_0 \\ z_0 \\ 0 \\ z_0 \end{bmatrix} .
\end{equation}
Then, for the second numerical equation (Eqs.~(\ref{equ:sys1}) to~(\ref{equ:sys5})) following the ray path, we use the following initial state:
\begin{equation}
   \begin{bmatrix}h \\ z \\ \theta \\ \zeta \\ b \end{bmatrix}_{h = H_\ttt{max}} = \begin{bmatrix}H_\ttt{max} \\ z_\infty \\ \theta_\infty \\ \zeta_\infty \\ 0 \end{bmatrix},
\end{equation}
where the values of $ z_\infty $, $ \theta_\infty,$ and $ \zeta_\infty$ are deduced from the previous computation, as these variables are part of the final state of the first equation system. 

While we previously chose $s$ the distance along the optical path as the integration variable, we have to define the integration limit as a function of the maximum height of Earth's atmosphere $H_\ttt{max}$: the altitude at which the density of air can be regarded as insignificant. The value of $H_\ttt{max}$ is often chosen equal to $80~\mrm{km}$ in the literature \citep{hohenkerk_computation_1985,van_der_werf_ray_2003}.

We relate the integration limit in $h$ and $s$ using the following approximation: 
\begin{equation}\label{equ:Smax}
    S_\ttt{max} = \frac{H_\ttt{max}}{\cos{z_0}}.
\end{equation}
This expression is not appropriate close to the horizon since the refraction angle $R_\text{a}$ grows and the approximation of ${1}/{\cos{\zeta}}$ in Eq.~(\ref{equ:sys1}) requires higher orders. However, this value of $S_\ttt{max}$ is an upper bound for the effective path length of the ray in the atmosphere of thickness $H_\ttt{max}$.

We also solve each system separately in the troposphere and in the stratosphere because of the discontinuity of the refractive index derivative at the tropopause (due to the discontinuity of the temperature gradient). However, it is not straightforward to separate the two integrations when using the variable $s$, as we do not know where the tropopause is located in the optical path. The idea is then to modify the last integration step that straddles the two domains in such a way that it ends at the tropopause. After that, we restart the integration with the initial constant step at the tropopause.

\subsection{Numerical accuracy of \texorpdfstring{$b^{(\infty)}$}{Lg} }\label{subsec:integration_step}
Due to numerical integration, the accuracy of the estimator $b^{(\infty)}$ necessarily depends on the chosen integration step. In order to set an integration step $\mrm{d}s$, we compute the angle of refraction for several steps starting from $\mrm{d}s = 2048~\mrm{m}$. We chose to carry out this study in terms of the refraction angle because this is the output of the integration of the first system (Eqs.~(\ref{equ:sys1}) to~(\ref{equ:sys4})). {The integration step is divided by two at each iteration until it equals the smallest reference step chosen at $\mathrm{d}s_\ttt{min}=0.5\,\ttt{m}$.} We choose such a small  {integration step} in order to have a reference that is affected by the error of the numerical integration as little as possible. 

\begin{figure}[!ht]
    \centering
    \includegraphics[width=0.45\textwidth,trim={0.8cm 0.2cm 0.2cm 0.8cm},clip]{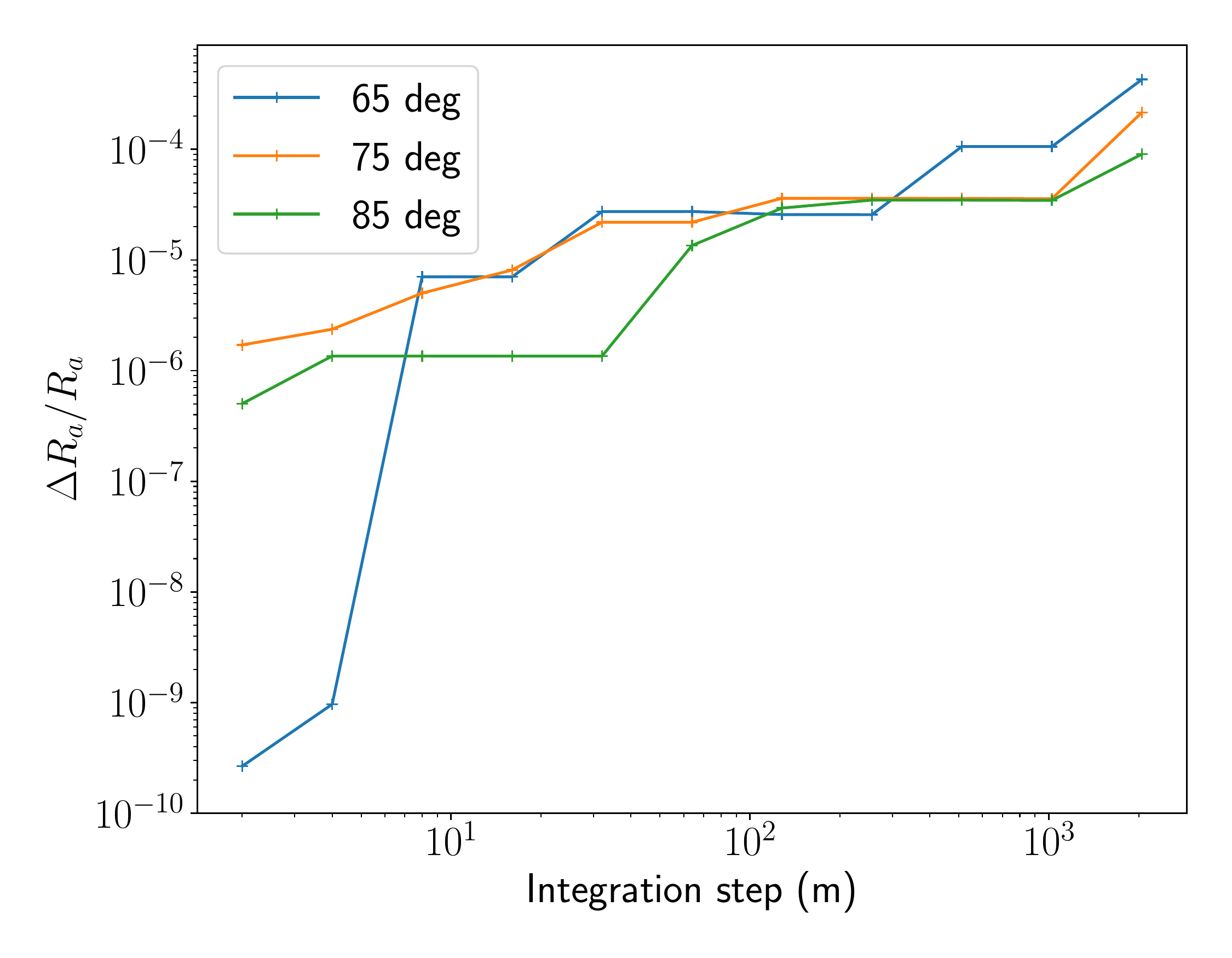}
    \caption{$\Delta R_\ttt{a}/R_\ttt{a}$ versus the integration step.}   
    \label{fig:b_integration_gap}
\end{figure}

The difference between the refraction angle at each $\mrm{d}s$ and its value for the smallest integration step $\mrm{d}s_\ttt{min}$ is plotted in Fig.~\ref{fig:b_integration_gap}:
\begin{equation}
    \Delta R_\ttt{a}/R_\ttt{a} = \left| \frac{R_\ttt{a}(\mrm{d}s) - R_\ttt{a}(\mrm{d}s_\ttt{min})}{R_\ttt{a}(\mrm{d}s_\ttt{min})} \right|, 
\end{equation}
and at three different zenith angles: $65^\circ$, $75^\circ,$ and $85^\circ$.

\begin{figure}[!ht]
    \centering
    \includegraphics[width=0.45\textwidth,trim={0.8cm 0.2cm 0.2cm 0.8cm},clip]{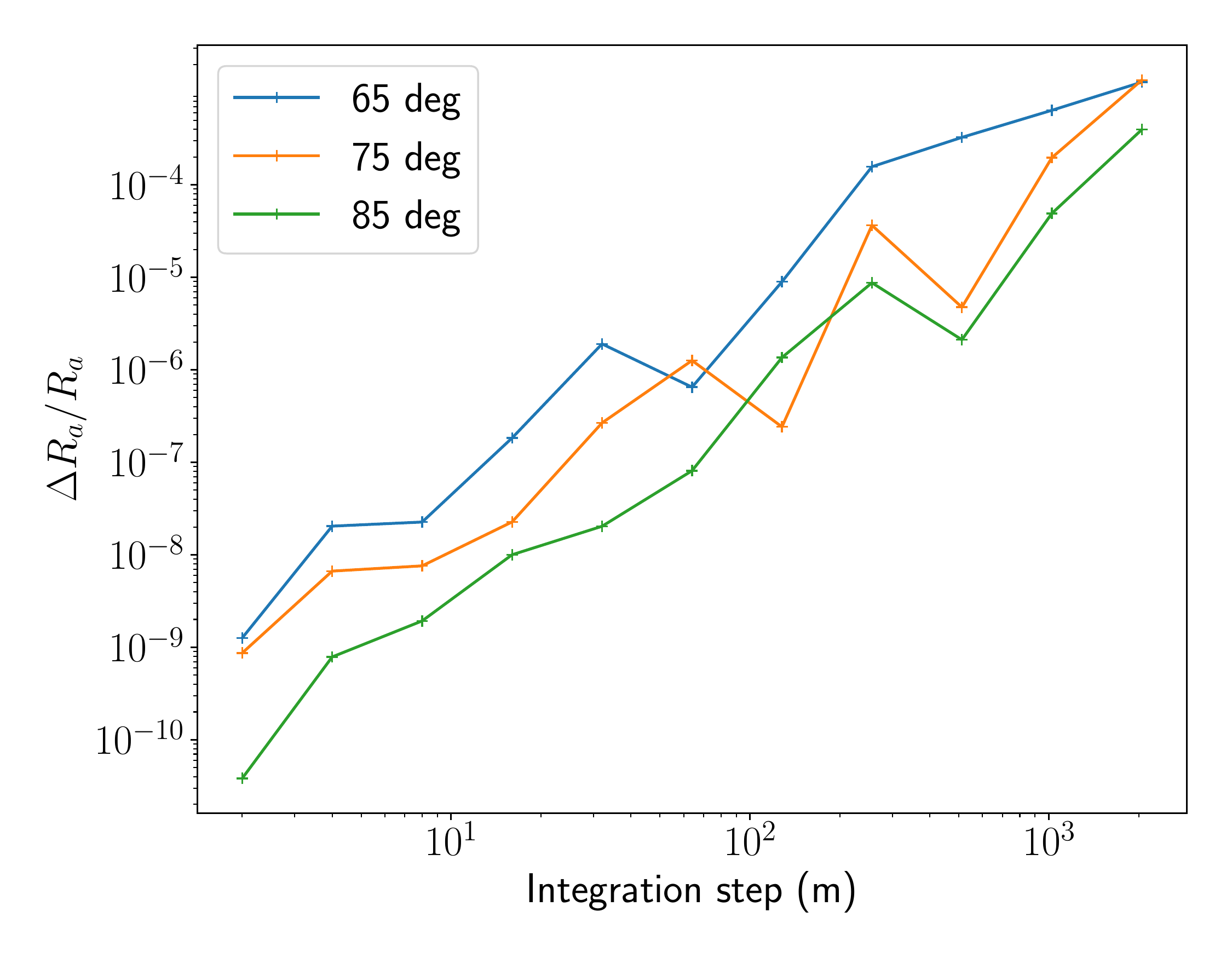}
    \caption{$\Delta R_\ttt{a}/R_\ttt{a}$ versus the integration step using a continuous derivative of the refractive index.}   
    \label{fig:b_integration_gap_modif}
\end{figure}

We notice in Fig.~\ref{fig:b_integration_gap} that the relative error decreases as the integration step $\mrm{d}s$ decreases. {However, the three curves often cross each other and do not decrease regularly, so some step changes have almost no effect on the value of the refraction angle. We suspect such irregularities to arise from the discontinuity of the refractive index derivative at the tropopause. We thus choose to recompute the error with a continuous profile of the refractive index derivative, using a second-order polynomial evolution of temperature near the tropopause, over 15 metres on each side. The result is shown in Fig.~\ref{fig:b_integration_gap_modif}. We can see that the relative error decreases globally while oscillating around a mean slope. The relative error is smaller for larger zenith angles, which is consistent with how $R_a$ increases with $z_0$. If we compare the two different models for an integration step $\mrm{d} s = 100\,\ttt{m}$, we find that the refraction angles vary by less than $20\,\ttt{milliarcseconds}$ at all zenith angles up to $85^\circ$. We therefore choose to settle for the discontinuous model. We also choose an integration step of $100~\mrm{m}$ as the relative error is smaller than $0.01\%$, which is, \textit{\emph{a priori,}} sufficient. } 

\subsection{Values of the lateral shift}
\begin{figure}[!ht]
    \centering
    \includegraphics[width=0.49\textwidth,trim={0cm 0cm 0cm 0cm},clip]{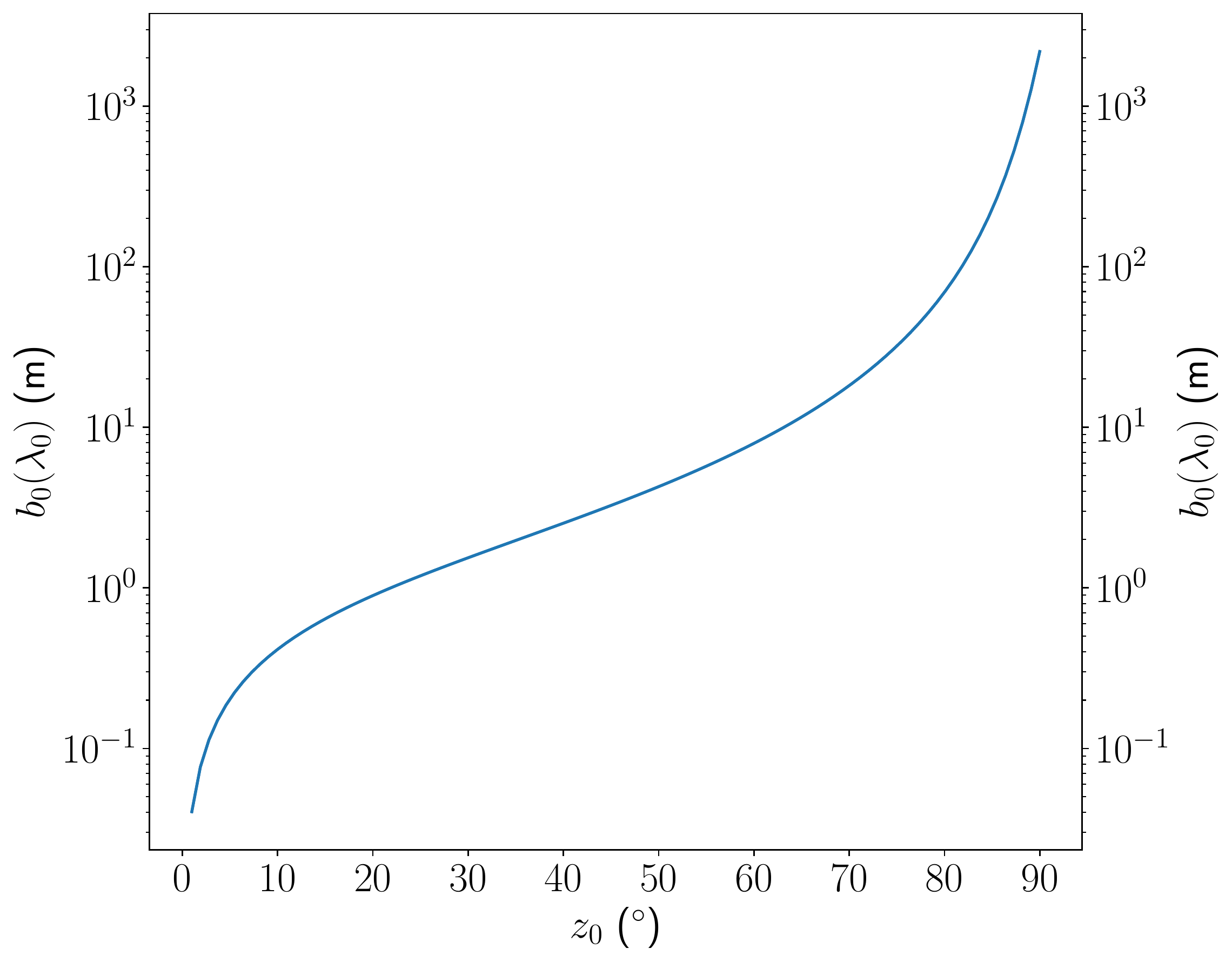}
    \caption{Lateral shift at $\lambda_0 = 550\, \mrm{nm}$ versus the apparent zenith angle, computed at SCTP using the estimator $b^{(\infty)}$.}
    \label{fig:sctp_shift}
\end{figure}

After dealing with the numerical aspects of the integration, here we study the evolution of the lateral shift using the Runge-Kutta estimator $b^{(\infty)}$ computed using the numerical method described in Sect. \ref{sec:numerical_method}. We start by plotting in Fig.~\ref{fig:sctp_shift} the evolution of the lateral shift at ground level as a function of the apparent zenith angle. The results are computed for the standard conditions of temperature and pressure (SCTP): 
\begin{equation}\label{equ:SCTP}
    T_\ttt{SCTP} = 273.15~\mrm{K} \text{, } P_\ttt{SCTP} = 1000~\mrm{hPa}.
\end{equation}
As a refe\-rence value for the wavelength $\lambda_0 = 550\, \mrm{nm}$, we chose the middle of the V spectral band.
We can see that the lateral shift increases with the apparent zenith angle; it tends towards {$0\,\mrm{m}$} when $z_0$ tends towards {$0^\circ,$} and it increases with a very steep slope near $z_0 = 90^\circ$, which matches perfectly with intuition and the evolution of the refraction angle $R_\ttt{a}$. The greater the zenith angle, the greater the angle of refraction and the greater the ray path through the atmosphere. We also note that the lateral shift reaches significant values beyond $20^\circ$ ($b\geq 1 ~\mrm{m}$), and can exceed tens of meters beyond $70^\circ$. For example, it equals {$3.3\,\ttt{m}$} at $z_0 = 45^\circ$ and is larger than $2\, \ttt{km}$ at $z_0 = 90^\circ$. 

\subsection{Influence of temperature and pressure}

\begin{figure}[!ht]
    \centering
    \includegraphics[width=0.49\textwidth,trim={0cm 0cm 0cm 0cm},clip]{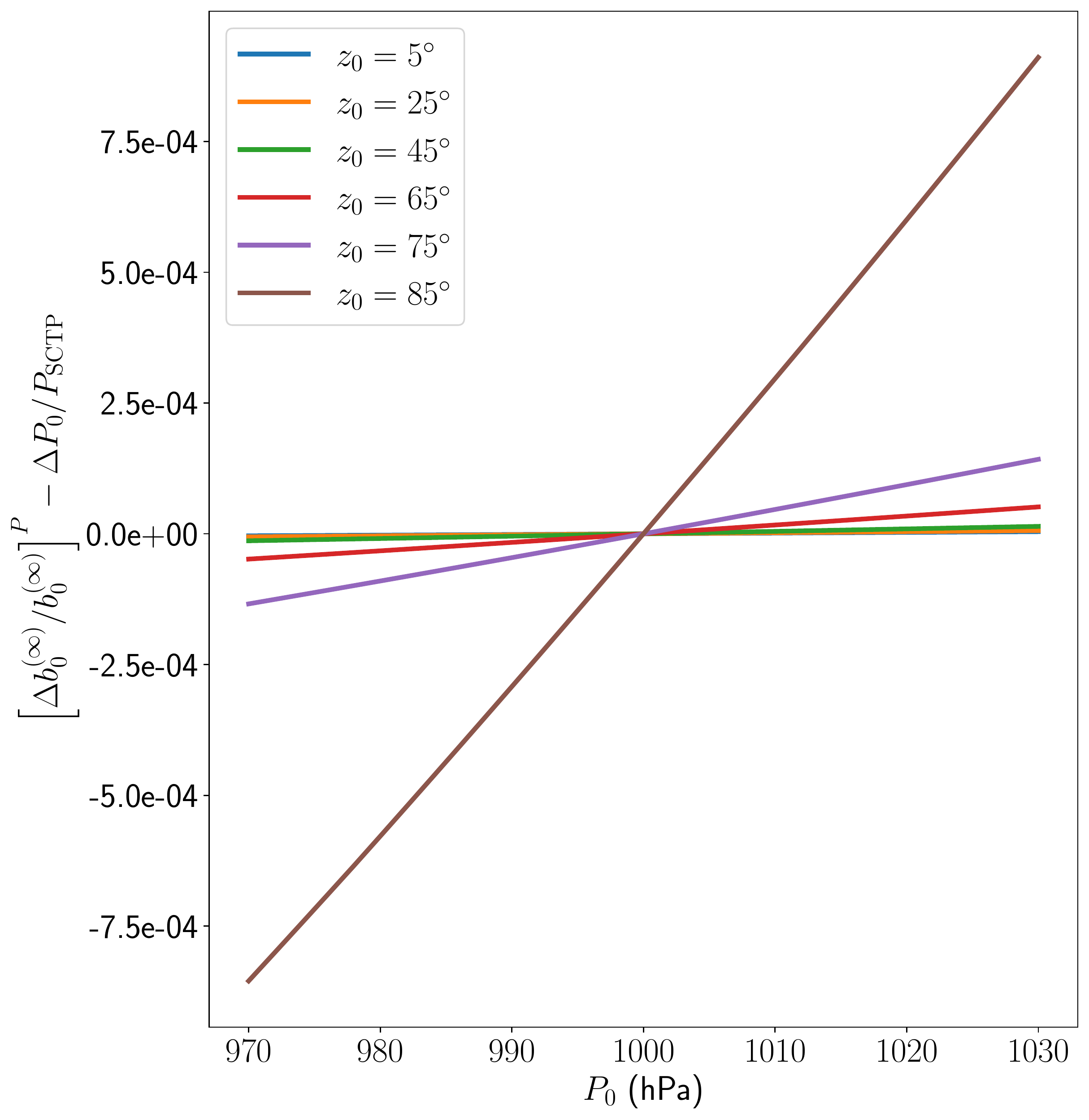}
    \caption{Deviation of $\left[{\Delta b_0}/{b_0}\right]^\ttt{P}$ at $\lambda_0$ from $\Delta P_0/P_\ttt{SCTP}$ versus ground pressure.}
    \label{fig:b_pressure_diff1}
\end{figure}

We now take a closer look at the distinct effects of pressure and temperature. We first deal with the pressure at the observer level, and we denote the relative difference between the lateral shift computed at a pressure $P_0$ and at the standard value $P_\ttt{SCTP} = 1000\,\ttt{hPa by }\left[
\Delta b_0 /b_0 \right]^\ttt{P}$, at $T_0 = T_\ttt{SCTP}$:
\begin{equation}
    \left[\frac{\Delta b_0}{b_0}\right]^\ttt{P} (P_0) = \frac{b_0(P_0) - b_0(P_\ttt{SCTP})}{b_0(P_\ttt{SCTP})}.
\end{equation}
The shift $b$ is a priori proportional to the amount of air above the observer, as confirmed by Eq.~(\ref{equ:b_wallner}) from \cite{wallner_effects_1976}, which shows that $b$ varies linearly with the pressure $P_0$ at the observer level. We can thus presume that
\begin{equation}\label{equ:b_pressure}
    \left[\frac{\Delta b_0}{b_0}\right]^\ttt{P}(P_0) \simeq \frac{P_0-P_\ttt{SCTP}}{P_\ttt{SCTP}}.
\end{equation}
This is confirmed by Fig.~\ref{fig:b_pressure_diff1} which shows that the deviation of $\left(\Delta b_0/b_0\right)(P_0)$ from its expected value (Eq.~(\ref{equ:b_pressure})) does not exceed $0.1\,\%,$ even for large zenith angles. In Fig.~\ref{fig:b_pressure_diff1}, we can also see that the residual increases with the zenith angle, but it still varies slightly with pressure. Moreover, the overall lateral shift increases with the pressure since the coefficient relating the lateral shift with pressure in Eq.~(\ref{equ:b_wallner}) is positive. 

\begin{figure}[!ht]
    \centering
    \includegraphics[width=0.49\textwidth,trim={0cm 0cm 0cm 0cm},clip]{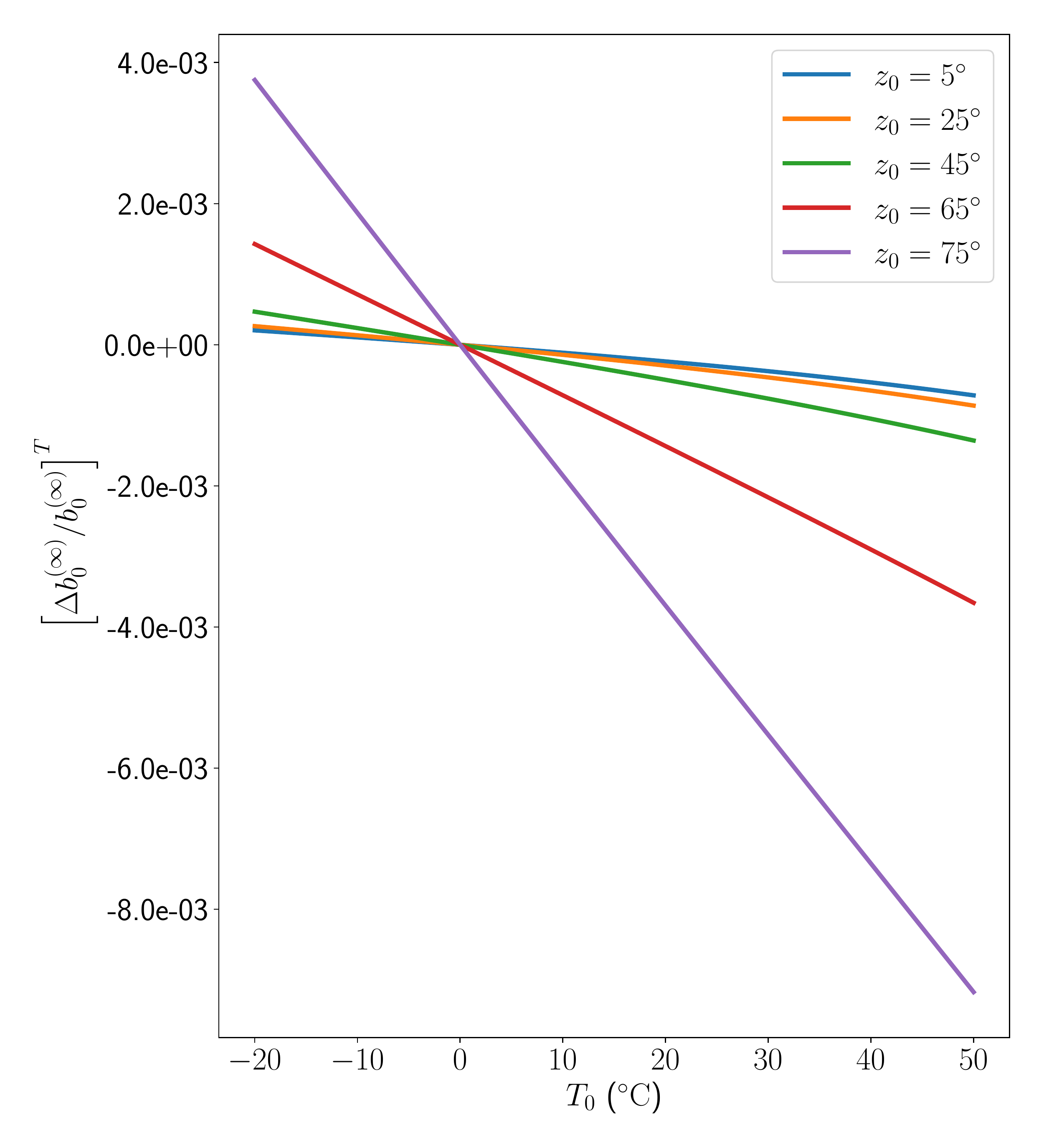}
    \caption{Variation of $\left[{\Delta b_0}/{b_0}\right]^\ttt{T}$ at $\lambda_0$ versus ground temperature.}
    \label{fig:b_temperature}
\end{figure}

We next study the effect of the temperature at the observer level. From Eq.~(\ref{equ:b_wallner}), we expect the effect of temperature to be negligible, and this is validated by Fig.~\ref{fig:b_temperature}, where we plot the relative difference between the lateral shift computed at a temperature $T_0$ and at the standard value $T_\ttt{SCTP} = 273.15\,\ttt{K}$ at $P_0 = P_\ttt{SCTP}$:
\begin{equation}
    \left[\frac{\Delta b_0}{b_0}\right]^\ttt{T}(T_0) = \frac{b_0(T_0) - b_0(T_\ttt{SCTP})}{b_0(T_\ttt{SCTP})},
\end{equation}
according to the temperature at the observer level $T_0$ and for several apparent zenith angles. For instance, we see in Fig.~\ref{fig:b_temperature} that the lateral shift decreases with temperature but its relative variation does not go above $1\%$. The greater the apparent zenith angle, the more visible the effects of temperature. We must also note that the two studied variations are in line with our expectations, since air density increases with pressure and decreases with temperature.

\subsection{Dependence on observation wavelength}
Since the refractive index depends on the observation wavelength $\lambda$, the effects of refraction, and, more specifically, the refraction angle and the lateral shift, also depend on $\lambda$. The angular deviation results in a chromatic blur effect (illustrated in Fig.~\ref{fig:refraction_intro}) that appears when observing a celestial body, since each light ray at wavelength $\lambda$ is bent by a different refraction angle $R_a(\lambda)$, while the studied shift induces a lateral translation of each ray path depending on the wavelength. Both issues are more pronounced when looking towards the horizon. In this section, we study the evolution of the lateral shift as a function of wavelength. 

\begin{figure}[!ht]
    \centering
    \includegraphics[width=0.49\textwidth,trim={0cm 0cm 0cm 0cm},clip]{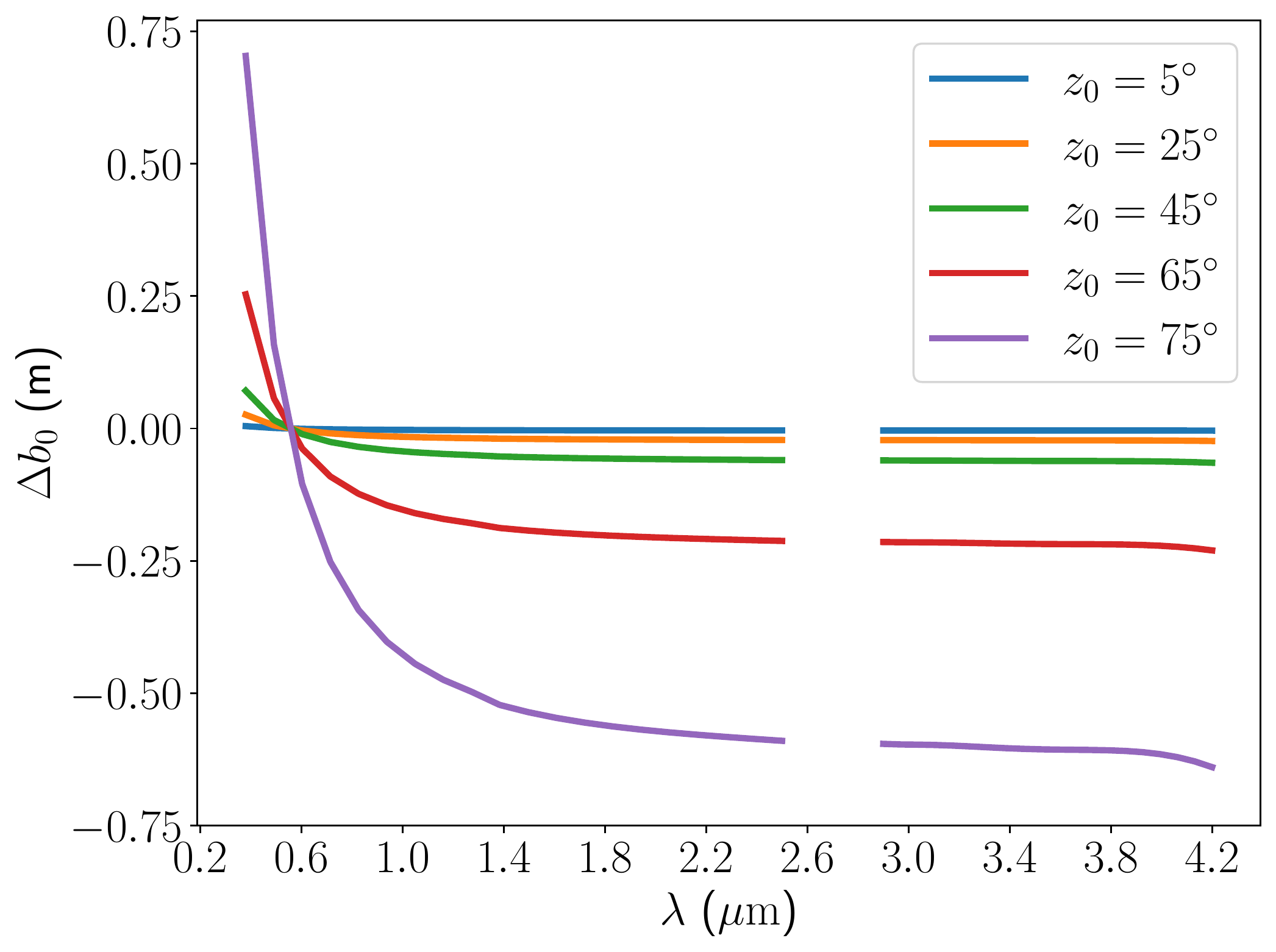}
    \caption{Absolute deviation of lateral shift from its value at $\lambda_0 = 550\,\mathrm{nm}$ as a function of wavelength, computed using the estimator $b^{(\infty)}$.}
    \label{fig:b_wavelength_absolu}
\end{figure}

In Fig.~\ref{fig:b_wavelength_absolu}, we plot the lateral shift at $h_0 = 0$ as a function of wavelength and for several apparent zenith angles:  
\begin{equation}
    {\Delta b}_0(z_0, \lambda) = b_0(z_0, \lambda) - b_0(z_0, \lambda_0).
\end{equation}

The results are computed for the SCTP without humidity and over a range of wavelengths from $0.4 \, \mrm{\mu m}$ to $4.2 \, \mrm{\mu m}$. Since Ciddor's refractivity formula \citep{ciddor_refractive_1996} is restricted to a maximum wavelength of $1.49 \, \mrm{\mu m}$, we covered the remaining interval using \cite{mathar_refractive_2007} formulas. {The slight jump in the curve of Fig.~\ref{fig:b_wavelength_absolu} is due to the transition between Ciddor's and Mathar's models at $\lambda = 1.3\,\mathrm{\mu m}$, while the gap between $2.5\,\mathrm{\mu m}$ and $2.9\,\mathrm{\mu m}$ results from the non-validity of Mathar's model in this region, due to high atmospheric absorption coefficients.}

\begin{figure}[!ht]
    \centering
    \includegraphics[width=0.49\textwidth,trim={0cm 0cm 0cm 0cm},clip]{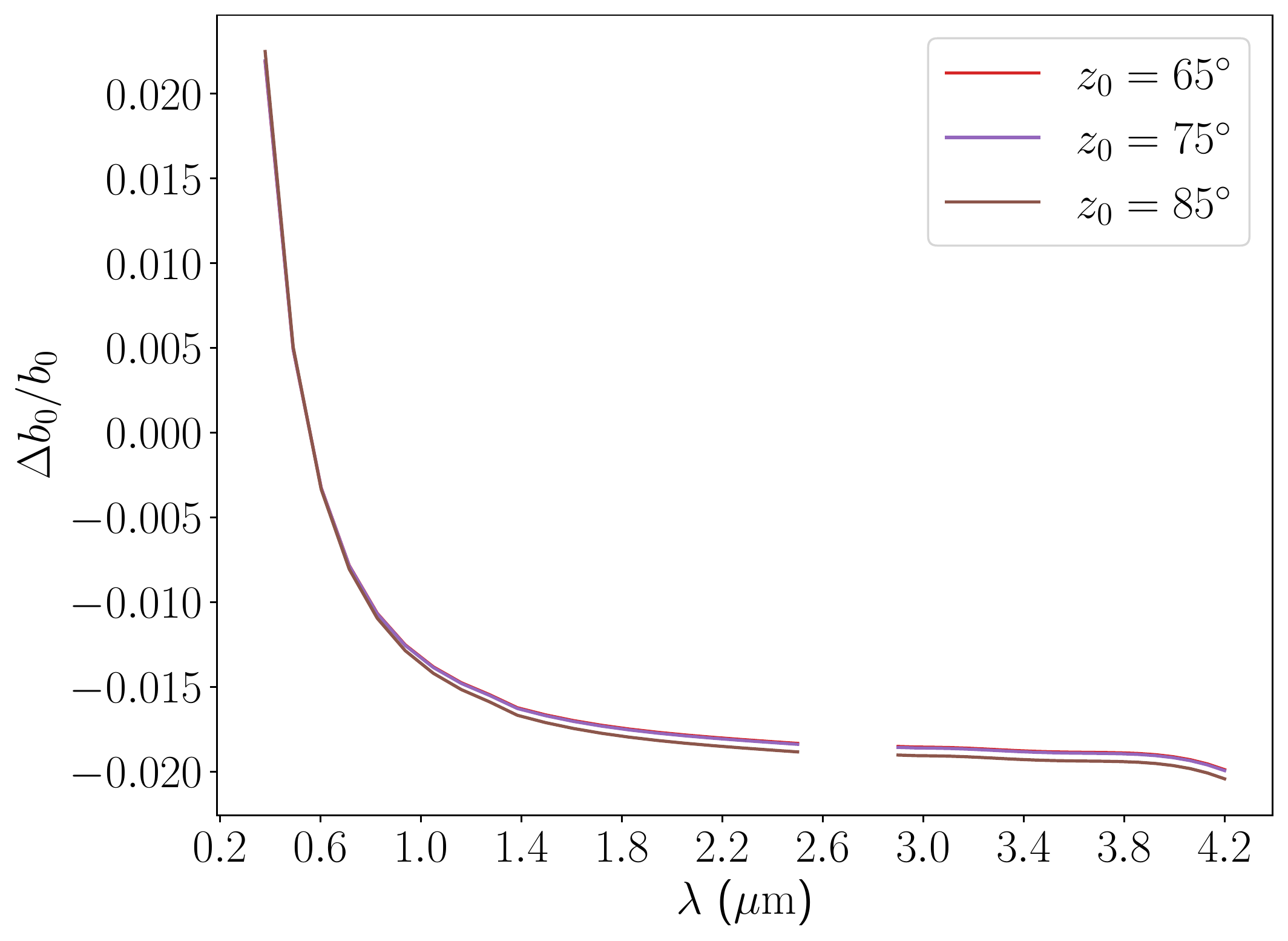}
    \caption{Relative deviation of lateral shift from its value at $\lambda_0 = 550\,\mathrm{nm}$ as a function of wavelength, computed using the estimator $b^{(\infty)}$.}
    \label{fig:b_wavelength_relatif}
\end{figure}

We observe that the lateral shift decreases with the wavelength for any zenith angle. In order to quantify the relative extent of this decrease, in Fig.~\ref{fig:b_wavelength_relatif} we plot the relative difference in $b(\lambda)$ with respect to its value at $\lambda_0$ and for a apparent zenith angles $z_0 = 65^\circ , 75^\circ, \text{ and } 85^\circ$:
\begin{equation}
    \frac{\Delta b_0}{b_0} (z_0, \lambda) = \frac{b_0(z_0, \lambda) - b_0(z_0, \lambda_0)}{ b_0(z_0, \lambda_0)} .
\end{equation}

We see that the variation of the lateral shift depending on the observation wavelength reaches at most $2\%$ of its value at $\lambda_0$ in the soft UV domain. We observe that the decay slope of the lateral shift also decreases along with the wavelength, until it becomes almost horizontal in the near- and mid-infrared.

We can also notice in Fig.~\ref{fig:b_wavelength_relatif} that the curves of $\Delta b_0/b_0$ are almost identical for the three zenith angles considered. This means that the lateral shift can be approximated by a function separable into two functions of the variables $\lambda$ and $z_0$; we see that this is indeed the case in the next section.

\section{Closed-form estimators of the lateral shift}\label{sec:taylor_expansion}
\subsection{Taylor expansion of the lateral shift}\label{sec:approx_2}

The system we obtained in Sect. \ref{subsec:equation_integration} \mLE{does not seem to have an analytical solution}, nor does it\rLE{ I removed 'allow us to'} reflect the influence of the main parameters (Earth's roundness, air density distribution, temperature and pressure at the observer) on the value of the lateral shift. In order to clarify this, we approximate the value of the lateral shift $b$ in a way similar to Laplace's formula for the refraction angle \citep{radau_recherches_1882}. Laplace's formula refers to an expression that allows the refraction angle to be calculated only from the apparent zenith angle and the weather conditions at sea level; it states that
\begin{align}\label{equ:laplace_formula}
    R_\ttt{a} &\simeq \alpha_0 \left( 1-\beta_0 \right) \, \tan{z_0} - \alpha_0 \left( \beta_0 - \frac{\alpha_0}{2} \right) \, \tan^3{z_0},
\intertext{where $\alpha_0$ and $\beta_0$ are defined by}
    \alpha_0 &= A_\ttt{D}(\lambda)\, \frac{P_0}{T_0},\label{equ:alpha_def}\\
    \beta_0 &= P_0/(h_0+R_\ttt{T})\rho_0, \label{equ:beta_def}
\end{align}
where $P_0$ and $T_0$ are, respectively, the temperature and pressure at the observation site, and $A_\ttt{D}$ is the refractivity of dry air defined in Eq.~(\ref{equ:Ad_ciddor}) ($\alpha_0$ still depends on the observation wavelength).

To begin with, we take advantage of the following: 
\begin{align}
     h &\ll R_\ttt{T}, \\
     n - 1 &\ll 1, \text{ and} \label{equ:n_negligeable} \\
     \sin{\left( z - z_\infty \right)} &\simeq z - z_\infty \label{equ:approx_sin_0}.
\end{align}

The three approximations are perfectly justified since they induce a total relative error lower than $1\,\%$. First, $h/R_\ttt{T}$ has as an upper bound $H_\ttt{max}/R_\ttt{pol} \approx 10^{-2}$ (where $R_\ttt{pol} =  6 356.75\,\ttt{km}$ is the polar radius of Earth, and $H_\ttt{max}$ is the height of the atmosphere defined in Eq.~(\ref{equ:Smax})). Secondly, the refractive index has its maximal value at ground level and does not exceed $10^{-3}+1$. Thirdly, the sine approximation is also valid, because the angle of refraction is always very small and does not exceed 30 arcminutes (at horizon). Even when considering real weather data, as in the study carried out by \cite{nauenberg_atmospheric_2017}, the error resulting from the sine approximation is smaller than $0.01 \%$.

Using the parameter $\alpha_0$ defined in Eq.~(\ref{equ:alpha_def}), the refractive index is written as follows:
\begin{equation}
    n = 1+ \alpha_0 \, \frac{T_0}{P_0} \, \frac{P}{T}.
\end{equation}
Since atmospheric air is assumed to behave according to the ideal gas law, one also obtains
\begin{equation}
    n = 1+\alpha_0 \, \Bar{\rho},
\end{equation}
where we denote, by $\Bar{\rho,}$ the normalised density $\rho / \rho_0$. Knowing Eqs.~(\ref{equ:alpha_def}) and~(\ref{equ:n_negligeable}), we have $\alpha_0 \ll 1$ as well.

In order to obtain the approximation, we start by integrating Eq.~(\ref{equ:equdb}) between $h_0$ (the altitude of the observation site) and the limit of Earth's atmosphere:
\begin{equation}
    b(h=h_0) - b(h\rightarrow \infty) =  \int^{h_0}_{\infty} \sin{ \left( z - z_{\infty} \right)} \,\mathrm{d}s.
\end{equation}

As depicted in Fig.~\ref{fig:devLat_notation_ds}, we choose to make the following calculations with $b(h\rightarrow \infty)=0$. 

With the objective of using the model of the atmosphere exposed in Appendix~\ref{app:dry_atmosphere}, we perform a change on the integration variable:
\begin{equation}\label{equ:b_approx_int}
     b(h=h_0) = \int^{h_0}_{\infty} \frac{\sin{\left( z - z_{\infty} \right)}}{\cos{\zeta}} \,\mathrm{d}h.
\end{equation}
We also restrict ourselves to the directions of observation sufficiently above the horizon to avoid divergences.

\begin{figure}[!ht]
    \centering
    \includegraphics[width=0.49\textwidth,trim={0cm 0cm 0cm 0cm},clip]{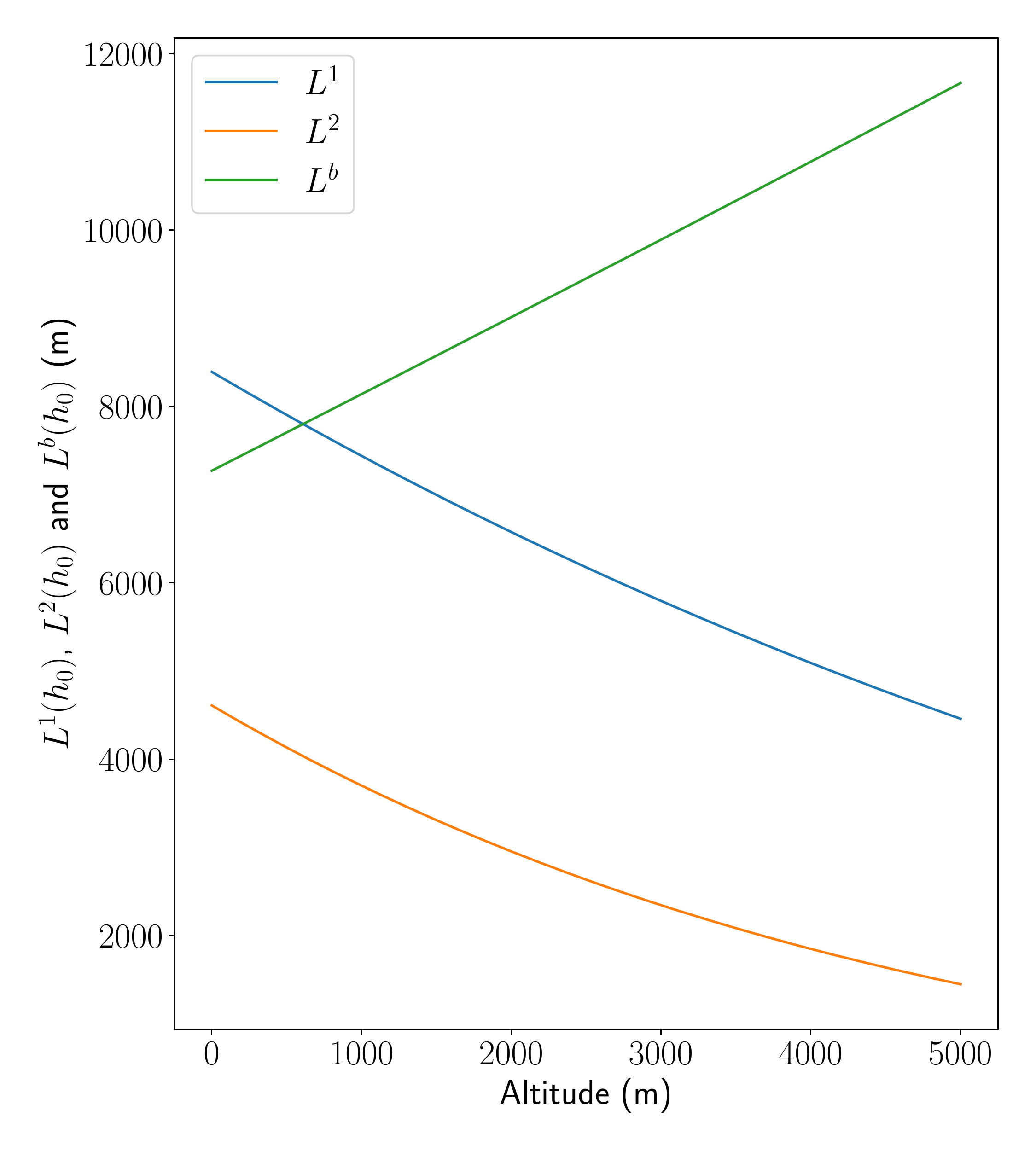}
    \caption{Evolution of the three characteristic lengths of the atmosphere as a function of the initial altitude.}
    \label{fig:longueur_approx2}
\end{figure}

Similarly to Laplace's formula, we aim to express the lateral shift in terms of the moments of the atmosphere. As the lateral shift's equation is more complex than the refraction angle equation, we define three moments:
\begin{subequations}
    \begin{align}\label{equ:moments_atm_0}
        L^1(h) & = \int_h^\infty \Bar{\rho} \, \mrm{d}x,\\
        L^2(h) & = \int_h^\infty \Bar{\rho}^2 \, \mrm{d}x,\\
        L^\ttt{b}(h)   & = \left[ \int_h^\infty x\,\Bar{\rho} \, \mrm{d}x \right] / L^1(h).
    \end{align}
\end{subequations}
We define $L^1_0$, $L^2_0$ and $L^\ttt{b}_0$ as $L^1(h_0)$, $L^2(h_0),$ and $L^\ttt{b}(h_0),$ respectively. They are explicitly calculated in Appendix~\ref{app:moment_atmosphere} as functions of temperature and pressure for the dry atmosphere model recalled in Appendix~\ref{app:dry_atmosphere}. Their evolution as a function of altitude is plotted in Fig.~\ref{fig:longueur_approx2}. 

To begin with, from Sect.~6.1.3 of \cite{kovalevsky_fundamentals_2004}, we obtain the following approximation of $z - z_\infty$: 
\begin{align}\label{equ:diff_z_zinfty}
    z(h) - z_{\infty} = & - \tan{z_0} \left( \alpha_0 \Bar{\rho} - \alpha_0^2 \Bar{\rho}^2 + \alpha_0^2 \Bar{\rho}  -\frac{h \alpha_0\Bar{\rho} }{R_\ttt{T} }- \frac{\alpha_0}{R_\ttt{T}} \int_h^\infty \Bar{\rho} \, \mrm{d}x \right) \notag \\
    & + \tan^3{z_0} \left(\alpha_0^2 \Bar{\rho} ^2 /2 - \alpha_0^2 \Bar{\rho} +\frac{h\alpha_0 \Bar{\rho} }{R_\ttt{T}} + \frac{\alpha_0}{R_\ttt{T}} \int_h^\infty \Bar{\rho} \, \mrm{d}x  \right).
\end{align}

Then, we expand the cosine term in Eq.~(\ref{equ:b_approx_int}) using Bouguer's invariant (Eq.~(\ref{equ:bouguer_invariant})):
\begin{subequations}
\begin{align}
    \frac{1}{\cos{\zeta}} & = \frac{1}{\sqrt{1-\sin^2{\zeta}}}, \\
    & = \frac{(h+R_\ttt{T})\, n / ( R_\ttt{T} n_0)}{ \sqrt{\left[\left(h+R_\ttt{T}\right)n/(R_\ttt{T}n_0)\right]^2-\sin^2{z_0}}}. \label{equ:cos_developpe}
\end{align}
\end{subequations}
In addition to this, since $\alpha_0 \ll 1$ and $h \ll R_\ttt{T}$, one obtains
\begin{subequations}
    \begin{align}
        \frac{(h+R_\ttt{T})\, n}{  R_\ttt{T} n_0} & = \left(1 + \frac{h}{R_\ttt{T}}\right) \left(\frac{1 + \alpha_0 \Bar{\rho}}{1+\alpha_0}\right),\\
        & = \left(1 + \frac{h}{R_\ttt{T}}\right) \left(1 + \alpha_0 \Bar{\rho}\right) \left(1 - \alpha_0 + o(\alpha_0) \right),\\
        & = 1 + \frac{h}{R_\ttt{T}} + \alpha_0 (\Bar{\rho} -1) + o\left(\alpha_0 + \frac{h}{R_\ttt{T}}\right).
    \end{align}
\end{subequations}
Based on this, we develop Eq.~(\ref{equ:cos_developpe}) to the second order with respect to $(h+R_\ttt{T})\, n/R_\ttt{T} n_0$. We obtain
   \begin{align}
        \frac{1}{\cos{\zeta}} = \frac{1}{\cos{z_0}}&-  \frac{\tan^2{z_0}}{\cos{z_0}}\left[\frac{(h+R_\ttt{T})\,n}{R_\ttt{T} \, n_0}-1\right]\notag \\
        & +\frac{3}{2}\left[ \frac{1}{\cos^5{z_0}}- \frac{1}{\cos^3{z_0}}\right]\left[\frac{(h+R_\ttt{T})\,n}{R_\ttt{T} \, n_0}-1\right]^2. \label{equ:approx_cos}
    \end{align}
As mentioned before, we note that this approximation is less accurate when the zenith angle $z_0$ approaches $90^\circ$ (i.e. when looking close to the horizon). The reason is that $1/\cos{z_0}$ is no longer defined when $z_0=90^\circ$. 

We then perform the product of the two approximations: Eqs.~(\ref{equ:diff_z_zinfty}) and~(\ref{equ:approx_cos}). This gives a second-order approximation $b^{(2)}$ of the lateral shift in a dry atmosphere:
\begin{align}\label{equ:approx_2_incomplete}
        b^{(2)} = & \frac{ \tan{z_0}}{\cos{z_0}} \int_{h_0}^{\infty} \left( \alpha_0 \Bar{\rho} - \alpha_0^2 \Bar{\rho}^2 + \alpha_0^2 \Bar{\rho} - \frac{\alpha_0 h\Bar{\rho}}{R_\ttt{T}} - \frac{\alpha_0}{R_\ttt{T}} \int_h^\infty \Bar{\rho} \, \mrm{d}x  \right) \mathrm{d}h\notag \\
       - &  \frac{ \tan^3{z_0}}{\cos{z_0}} \int_{h_0}^{\infty} \left( \frac32\alpha_0^2 \Bar{\rho}^2 - 2\alpha_0^2 \Bar{\rho} + \frac{2\alpha_0 h \Bar{\rho}}{R_\ttt{T}} + \frac{\alpha_0}{R_\ttt{T}}\int_h^\infty \Bar{\rho} \, \mrm{d}x  \right)  \mathrm{d}h .
\end{align}  

We can directly write most of the terms involved in Eq.~(\ref{equ:approx_2_incomplete}) using the three moments of the atmosphere, except for the last term, which nests two integrals. For the latter, we perform an integration by \mLE{parts}\rLE{the s was missing, it is a very common mathematical process}: 
\begin{equation}\label{equ:egalite_int}
    \int_{h_0}^{\infty} \int_{h}^{\infty} \Bar{\rho} \,\mathrm{d}x  \mathrm{d}h  = \int_{h_0}^{\infty} L^1(h) \, \mathrm{d}h  = -\frac{h_0 P_0}{g \rho_0}+ \int_{h_0}^{\infty}h  \Bar{\rho} \, \mathrm{d}h .     
\end{equation}

Lastly, using Eq.~(\ref{equ:egalite_int}), we write the lateral shift approximation using the three moments of the atmosphere. We obtain
\begin{equation}\label{equ:b2_approx}
    b^{(2)} = A_\ttt{sh}\frac{\tan{z_0}}{\cos{z_0}} - B_\ttt{sh} \frac{\tan^3{z_0}}{\cos{z_0}},
\end{equation}
where
\begin{align}\label{equ:Ashift}
    A_\ttt{sh} &= \alpha_0 \left(1+\frac{h_0}{R_\ttt{T}}\right)L^1_0 -2\alpha_0 L^\ttt{b}_0 \frac{L^1(h)}{R_\ttt{T}} - \alpha_0^2 \left( L^2_0 - L^1_0 \right),
   \\ \label{equ:Bshift}
    B_\ttt{sh} &= -\frac{\alpha_0 h_0}{R_\ttt{T}} L^1_0+ 3\alpha_0 L^\ttt{b}_0 \frac{L^1_0}{R_\ttt{T}} + \alpha_0^2 \left(\frac32 L^2_0 - 2 L^1_0 \right) .
\end{align}

The final expression in Eq.~(\ref{equ:b2_approx}) efficiently decouples the contribution of the atmosphere to the lateral shift from the contribution of the apparent zenith angle and the observation wavelength. It also spares the user having to integrate an integrand varying with the zenith angle. Thus, we have somehow established a Laplace formula for the lateral shift. Nevertheless, there is a major difference between the two formulas because of the two moments $L^2$ and $L^\ttt{b}$ in the expression of $b^{(2)}$.  Indeed, these two lengths depend on the composition of the atmosphere and the distribution of air within it. In the following section, we describe the terms that are part of our approximation in detail.

\subsection{First-order estimator: \texorpdfstring{$b^{(1)}$}{Lg}}
In the approximate analytic formula for the lateral shift established in Sect. \ref{sec:approx_2}, one can distinguish the first-order term that is only proportional to $\alpha_0$, which we call $b^{(1)}$:
\begin{equation}\label{equ:approx_shift_1}
    b^{(1)} (h_0,z_0) = \,\alpha_0 \, \frac{\tan{z_0}}{\cos{z_0}}L^1_0.
\end{equation}
We thus find here Eq.~(\ref{equ:b_wallner}) recalled in the introduction and which has been given in the literature by several authors. Yet, the difference between Eq.~(\ref{equ:b_wallner}\&\ref{equ:approx_shift_1}) is the angle used in the tangent and cosine terms. In Eq.~(\ref{equ:b_wallner}) there is a $z_\infty,$ while in Eq.~(\ref{equ:approx_shift_1}) it is $z_0$. Fortunately, these two equations are equivalent because the refraction angle $R_\ttt{a}$ is very small. Equation~(\ref{equ:b_wallner}) has been demonstrated by Wallner under the assumption of a flat Earth, as confirmed here, since the first-order part of $b^{(2)}$ tends towards $b^{(1)}$ when $R_\ttt{T}$ tends towards $\infty$. 

\subsection{Second-order estimator: \texorpdfstring{$b^{(2)}$}{Lg}}
Next, we take a closer look at the second-order approximation of the lateral shift Eq.~(\ref{equ:b2_approx}). We can distinguish two different terms \mLE{whose approximation orders are} higher than \mLE{the first-order}\rLE{ the two terms are not strictly speaking higher than one, they approximate the lateral shift to an order higher than the first-order}: \mLE{a term $b^{(2)}_\ttt{R}$ proportional to $\alpha_0/R_\ttt{T,}$ which is due to Earth's roundness; and another term, $b^{(2)}_2,$ proportional to $\alpha_0^2$.}

We have:
\begin{equation}
    b^{(2)} = b^{(1)} + b^{(2)}_\ttt{R} + b^{(2)}_2 
,\end{equation}
where 
\begin{subequations}
    \begin{align}
        b^{(2)}_\ttt{R} (h_0) =& \frac{\alpha_0}{R_\ttt{T}}\frac{\tan{z_0}}{\cos^3{z_0}}L^1_0 \left[h_0 - 3 L^{b}_0\right] 
        +\frac{\alpha_0}{R_\ttt{T}}\frac{\tan{z_0}}{\cos{z_0}}L^1_0 L^\ttt{b}_0,\\
        b^{(2)}_2 (h_0) =& \, \alpha_0^2 \frac{\tan{z_0}}{\cos{z_0}} \left[ L^1_0 - L^2_0\right] 
        - \alpha_0^2 \frac{\tan^3{z_0}}{\cos{z_0}} \left[\frac{3}{2}L^2_0-2L^1_0 \right].
    \end{align}
\end{subequations}
We note that the Earth's radius $R_\ttt{T}$ does not appear in the second term $b^{(2)}_2$, so it is the only  second-order term within the approximation of a flat Earth. 
\begin{figure}[!ht]
    \centering
    \includegraphics[width=0.49\textwidth,trim={0.3cm 0cm 0cm 0cm},clip]{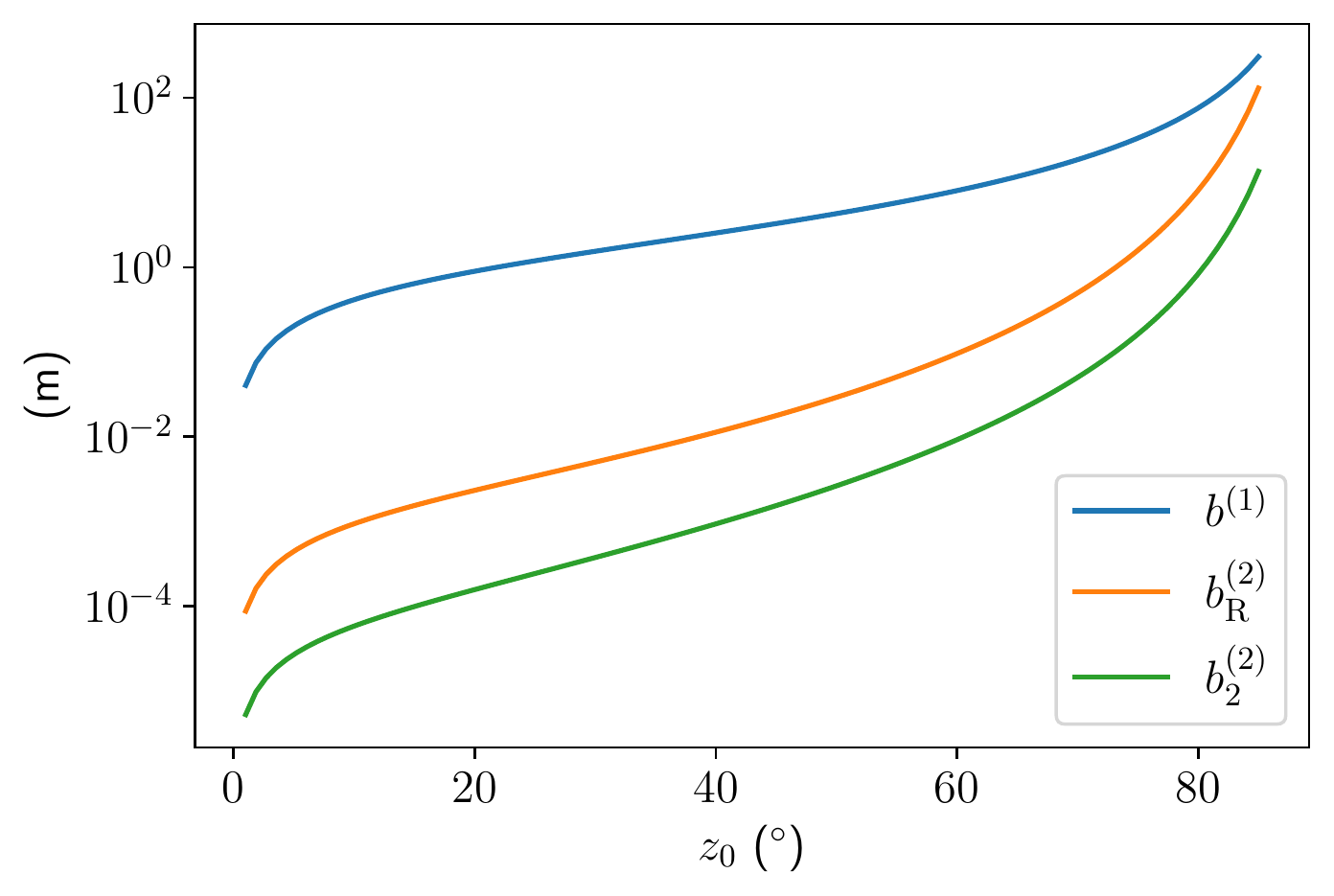}
    \caption{Evolution of the three terms in $b^{(2)}$ according to the zenith angle.}
    \label{fig:separation_bLaplace_rotondite_atm}
\end{figure}

In order to capture the influence of each term, in Fig.~\ref{fig:separation_bLaplace_rotondite_atm} we plot the evolution in absolute values of each of the three terms according to the zenith angle. We calculated the multiplying factor $\alpha_0$ at the wavelength $\lambda_0$. In Fig.~\ref{fig:separation_bLaplace_rotondite_atm}, we observe that the main contribution comes from the flat Earth term $b^{(1)}$ and is followed by Earth's roundness $b^{(2)}_R,$ and then comes the second-order term $b^{(2)}_2$.

\subsection{A separable third estimator: \texorpdfstring{$b^{(3/2)}$}{Lg}}
\begin{figure}[!ht]
    \centering
    \includegraphics[width=0.49\textwidth,trim={0cm 0cm 0cm 0cm},clip]{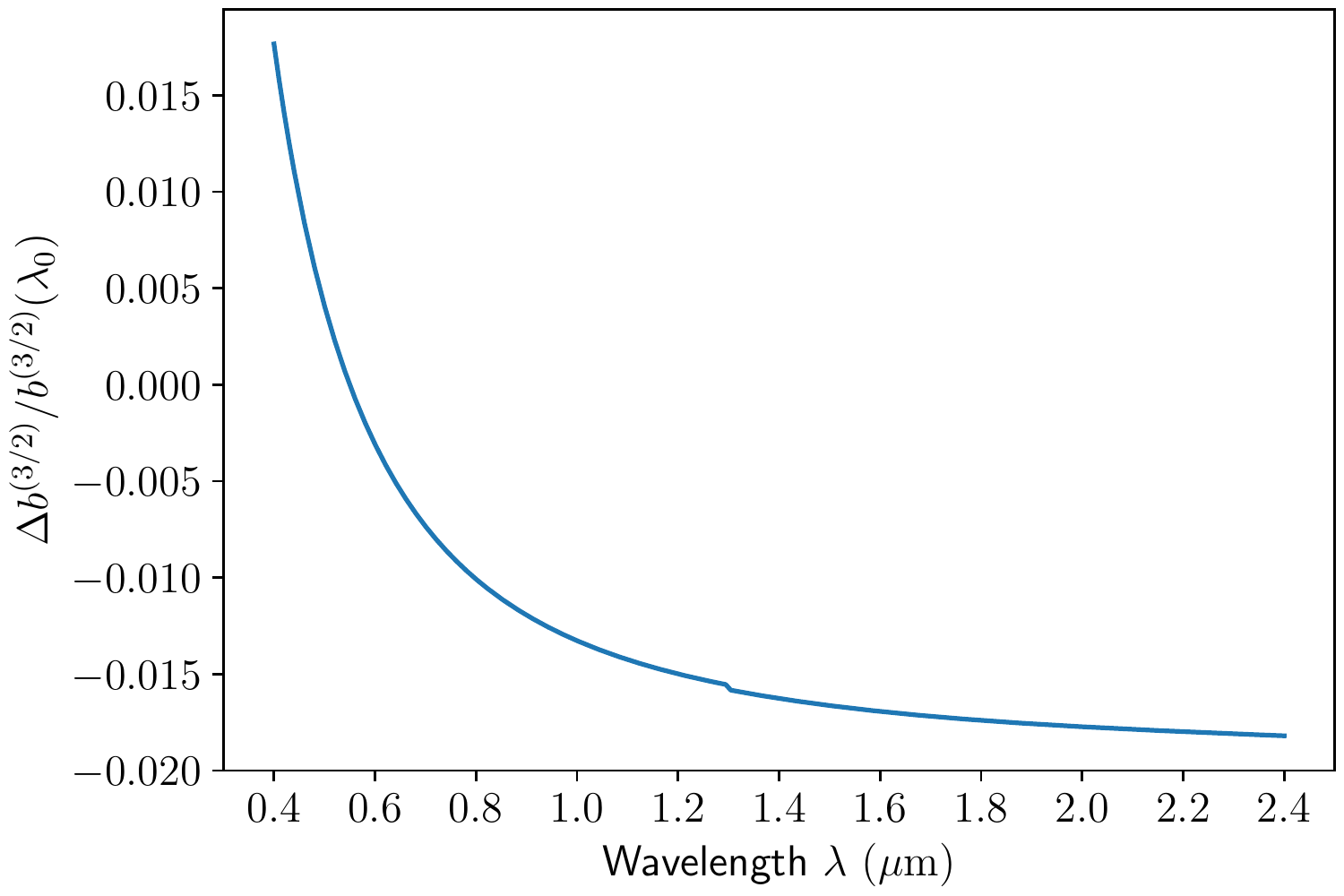}
    \caption{Chromatic behaviour of the normalised relative shift  $\Delta_r b^{(3/2)}$ with respect to $\lambda_0 = 550\,\ttt{nm}$ using the estimator $b^{(3/2)}$.}
    \label{fig:deltab32}
\end{figure}

The dominance of the term due to Earth's roundness $b_\ttt{R}^{(2)}$ over the term $b_2^{(2)}$ drives us to define a third estimator of the lateral shift denoted $b^{(3/2)}$, composed of the two terms $b^{(1)}$ and $b^{(2)}_\mrm{R}$, such that
\begin{equation}
    b^{(3/2)} = b^{(1)} + b^{(2)}_\mrm{R}.
\end{equation}
This third approximation, between the first and second orders, has the advantage of decoupling the wavelength (only in $\alpha_0$) from the zenith angle. Thanks to this, this approximation of the lateral shift is separable into functions of $\lambda$ and $z_0$:
\begin{align}\label{equ:b32}
   b^{(3/2)}(h_0,z_0) =& \alpha_0\frac{L^1_0}{R_\ttt{T}} \frac{\tan{z_0}}{\cos{z_0}} \left\{ R_\ttt{T} + L^\ttt{b}_0 + \frac{1}{\cos^2{z_0}} \left[ h_0 - 3 L^{b}(h_0)\right] \right\}.
\end{align}
In Fig.~\ref{fig:deltab32}, we plot the value of
\begin{equation}\label{equ:deltarB32}
    \Delta_r b^{(3/2)} (\lambda) = \frac{\Delta b^{(3/2)} (z_0, \lambda)}{b^{(3/2)}(z_0, \lambda_0)} = \frac{b^{(3/2)}(z_0, \lambda) - b^{(3/2)}(z_0, \lambda_0)}{b^{(3/2)}(z_0, \lambda_0)} 
\end{equation}
as a function of $\lambda$. This quantity depends only on the wavelength $\lambda$, and is in fact simply equal to 
\begin{equation}\label{equ:b32_n}
    \Delta_r b^{(3/2)} (\lambda) = \frac{\alpha_0 (\lambda) - \alpha_0 (\lambda_0)}{\alpha_0 (\lambda_0)} = \frac{n(\lambda) - n(\lambda_0)}{n(\lambda_0)} .
\end{equation}

The values of $\Delta_r b^{(3/2)}$ plotted in Fig.~\ref{fig:deltab32} are constant with respect to the zenith angle and can be directly computed with Eq.~\ref{equ:b32_n} using the refractivity from Eq.~(\ref{equ:Ad_ciddor}). Consequently, one can calculate the chromaticity of the lateral shift at any zenith angle $z$ simply by performing the product of $\Delta_r b^{(3/2)}$ with the lateral shift at the chosen zenith angle and at $\lambda_0$ ($b^{(\infty)}(z,\lambda_0)$).

\subsection{Approximation error of \texorpdfstring{$b^{(1)}$}{Lg}, \texorpdfstring{$b^{(2)}$}{Lg} , and \texorpdfstring{$b^{(3/2)}$}{Lg} with respect to \texorpdfstring{$b^{(\infty)}$}{Lg}}\label{sec:accuracy_estimators}
In order to quantify the accuracy of the previously established approximations, we compute the values of the three estimators $b^{(1)}$, $b^{(3/2)}$, and $b^{(2)}$ for various zenith angles and compare them to the Runge-Kutta estimator $b^{(\infty)}$ (all at the SCTP) at sea level and with an observation wavelength equal to $\lambda_0$.

\begin{figure}[!ht]
    \centering
    \includegraphics[width=0.49\textwidth,trim={0cm 0cm 0cm 0cm},clip]{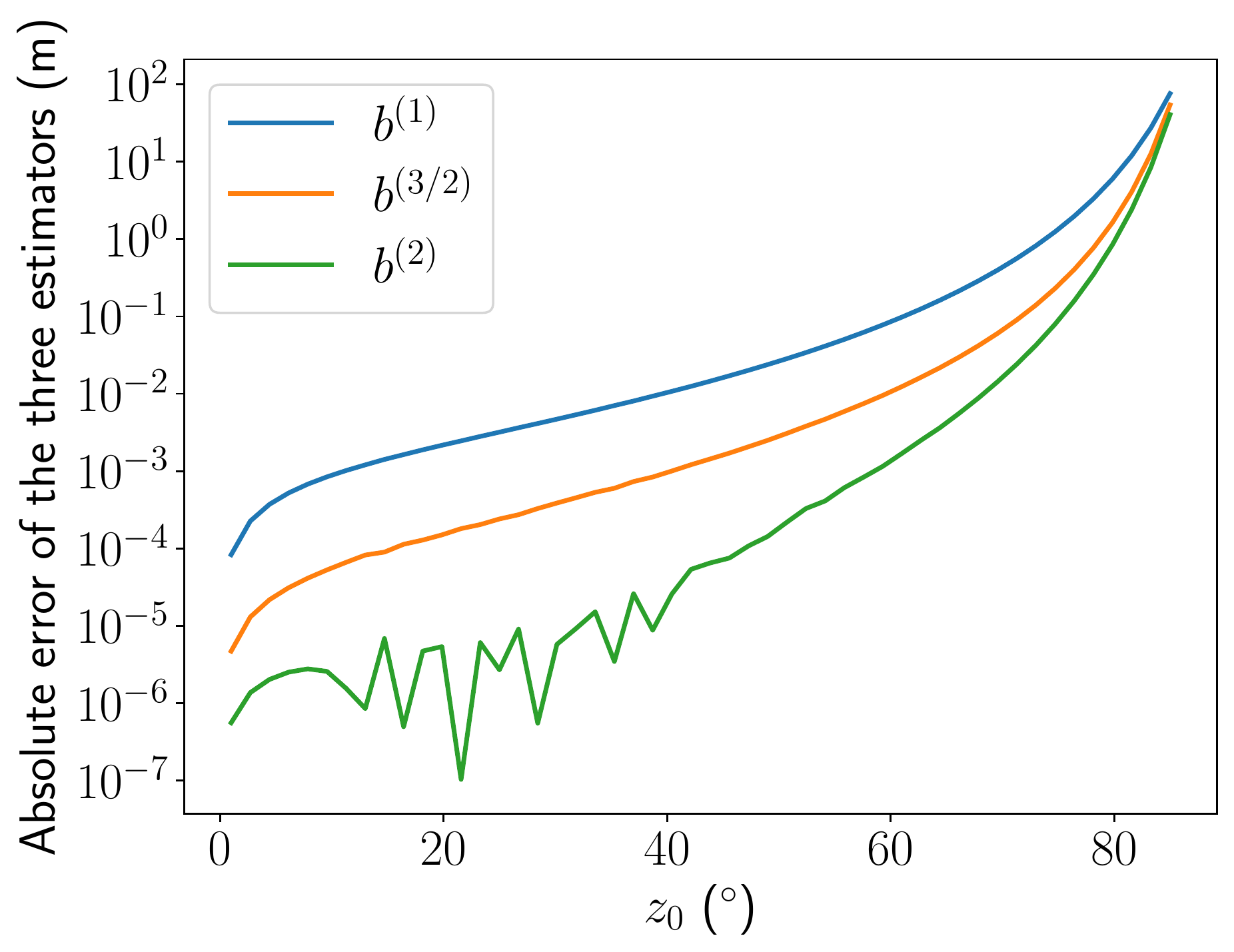}
    \caption{Absolute error of estimators $b^{(1)}$, $b^{(3/2)}$, and $b^{(2)}$ with respect to the Runge-Kutta estimator $b^{(\infty)}$.}
    \label{fig:ecart_b_laplace}
\end{figure}

\begin{figure}[!ht]
    \centering
    \includegraphics[width=0.49\textwidth,trim={0cm 0cm 0cm 0cm},clip]{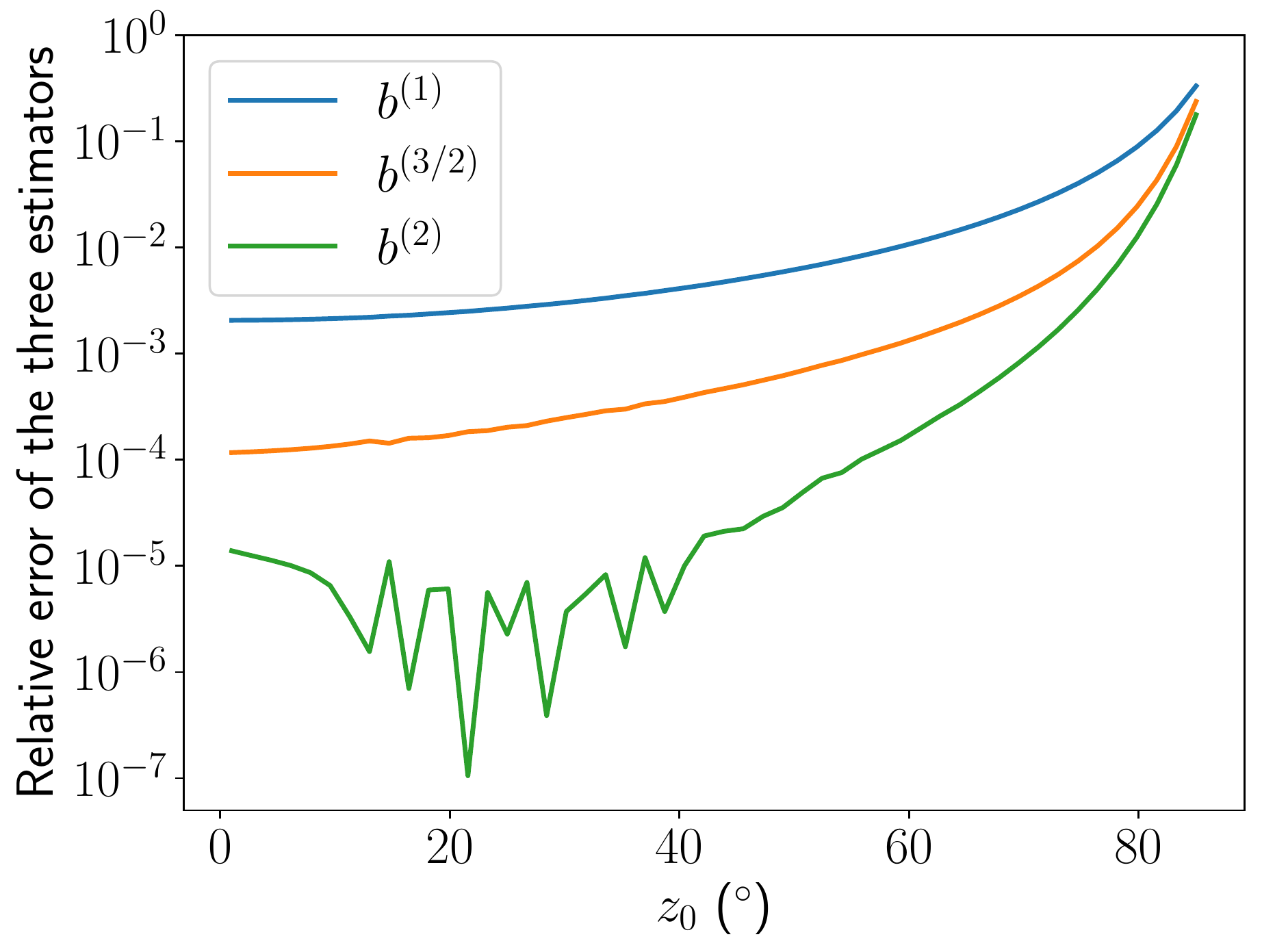}
    \caption{Relative error of the estimators $b^{(1)}$, $b^{(3/2)}$, and $b^{(2)}$ with respect to the Runge-Kutta estimator $b^{(\infty)}$.}
    \label{fig:ecart_relatif_b_laplace}
\end{figure}

We plot the absolute and relative errors of the three approximations of the lateral shift, respectively, in Figs. \ref{fig:ecart_b_laplace} and \ref{fig:ecart_relatif_b_laplace}: 
\begin{align}
    \epsilon^{(X)}_\ttt{a}(z_0) & = \left| b^{(X)}(h_0 = 0, z_0) - b^{(\infty)}(h_0 = 0,z_0) \right|, \\
    \epsilon^{(X)}_\ttt{r}(z_0) & = \left| \frac{b^{(X)}(h_0 = 0, z_0) - b^{(\infty)}(h_0 = 0,z_0)}{b^{(\infty)}(h_0 = 0,z_0)} \right| ,
\end{align}
where $X$ stands for $1$, $3/2,$ or $2$. The fluctuations observed in the curves of $b^{(2)}$ in Fig.~\ref{fig:ecart_b_laplace}~and~\ref{fig:ecart_relatif_b_laplace} are numerical and are due to the choice of the integration step for the estimator $b^{(\infty)}$. The chosen step ($100\,\ttt{m}$) allows us to make quick computations but is not small enough to ensure that the residual of the numerical integration is lower than $10^{-5} \,\ttt{m}$.

It is clear that the three approximations deviate from the numerical solution as the zenith angle increases. This is consistent with the approximation that has been made in Eq.~(\ref{equ:approx_cos}) and that is also implicitly present in Laplace's formula and therefore in Eq.~(\ref{equ:diff_z_zinfty}). From Fig.~\ref{fig:ecart_relatif_b_laplace}, we can see that the second-order approximation has a relative error of less than $1\%$  up to a zenith angle of $75^\circ,$ while the first-order approximation satisfies this requirement up to roughly $60^\circ$ only.  Concerning the approximation $b^{(3/2)}$, it is right in between the two 'complete' order approximations; it suffers ten~times less relative error than order 1 and thus achieves a relative error of less than 1\%  up to $z_0 = 70^\circ$. Specifically, the $b^{(3/2)}$ estimator achieves errors of less than 0.1\% up to a zenith angle of $50^\circ$. We also notice on Fig.~\ref{fig:ecart_relatif_b_laplace} that the relevance of high-order estimators is mainly found for zenith angles lower than 80$^\circ$. Indeed, the relative error curves of the three estimators are very close for $z_0 \geq 80^\circ$. Finally, the choice of the appropriate estimator will depend on the tolerated error levels and the kind of application.

\section{Applications} \label{sec:application}

\subsection{Wavefront sensing}\label{sec:lateral_shift_OA}
This section focuses on the impact of the lateral shift in the performance of wavefront sensing systems. When dealing with atmospheric turbulence, in order to improve photometric efficiency it is usual to use different wavelengths for the science channel (the wavelength of interest for the celestial object) and the wavefront sensing channel. Yet, this scheme induces different optical paths for the related ray because of the chromaticity of the lateral shift, and hence the phase distortions seen by the two channels differ. As indicated in \cite{wallner_effects_1976} and  \cite{nakajima_zenith-distance_2006}, the values of the lateral shift all along the ray path at the two wavelengths ($\lambda_\ttt{SCI}$ in the scientific channel and $\lambda_\ttt{WFS}$ in the wavefront sensing channel) allow us to compute the wavefront phase variance associated with the chromatic shear by the use of a specific and known turbulence profile. 

Our goal here is not to duplicate the mathematical developments present in the previously cited papers, but only to discuss the relevance of calculating the lateral displacement of the ray more precisely, \mLE{that is }using numerical computation instead of the first-order approximation. 

\begin{figure}[!ht]
    \centering
    \includegraphics[width=0.49\textwidth,trim={0cm 0cm 0cm 0cm},clip]{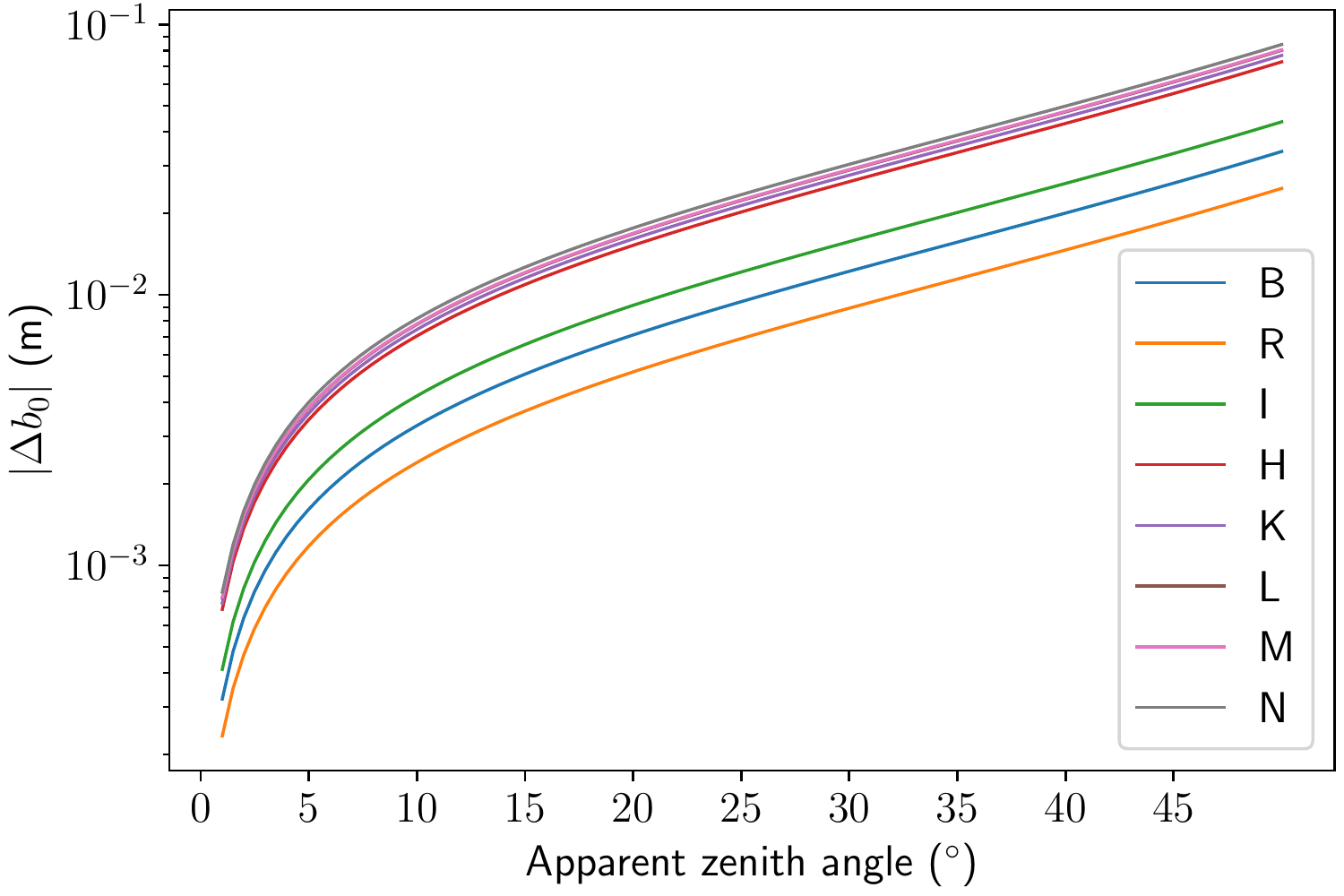}
    \caption{Difference between the lateral shift at $\lambda_\ttt{WFS} = 550\,\ttt{nm}$ and various $\lambda_\ttt{SCI}$ versus zenith angle, computed using the estimator $b^{(\infty)}$. }
    \label{fig:b_AO_diff}
\end{figure}

To begin with,  we assume $\lambda_\ttt{WFS} = 550 \,\ttt{nm}$ and plot on Fig.~\ref{fig:b_AO_diff} the difference between the value of the lateral shift at $\lambda_\ttt{WFS}$ and its value at $\lambda_\ttt{SCI}$ as a function of the observation zenith angle. Several values of $\lambda_\ttt{SCI}$ are used; these are the central wavelengths of the spectral bands B ($445\,\ttt{nm}$), R ($658\,\ttt{nm}$), I ($806 \,\ttt{nm}$), H ($1630\,\ttt{nm}$), K ($2190\,\ttt{nm}$), L ($3450\,\ttt{nm}$), M ($4750\,\ttt{nm}$), and N ($10500\,\ttt{nm}$). We define $\Delta b_0$ as 
\begin{equation}
    \Delta b_0  = b_0 (z_0, \lambda_\ttt{WFS}) - b_0 (z_0, \lambda_\ttt{SCI})
.\end{equation}

The lateral shift was purposely calculated over the entire thickness of the atmosphere, so as to simplify the computation and not consider any specific observation site. Still, this is a worst case scenario for the lateral shift as the astronomical observatories where wavefront sensing is used are located at altitudes around $3\,\mathrm{km}$ or above. In {Fig.~\ref{fig:b_AO_diff}, we observe} that the deviation $\Delta b_0$  increases both as functions of the zenith angle and the wavelength. The greater the difference between $\lambda_\ttt{WFS}$ and $\lambda_\ttt{SCI}$, the greater $\Delta b_0$, because of the monotonic nature of the lateral shift as a function of the wavelength. We also notice the stationarity that we saw in Fig.~\ref{fig:b_wavelength_absolu}, the curves of $\Delta b_0$ with a $\lambda_\ttt{SCI}$ greater than $1630\,\ttt{nm}$ are almost identical. We note that the B band is located between the R and I spectral bands due to the absolute value.

\begin{figure}[!ht]
    \centering
    \includegraphics[width=0.49\textwidth,trim={0cm 0cm 0cm 0cm},clip]{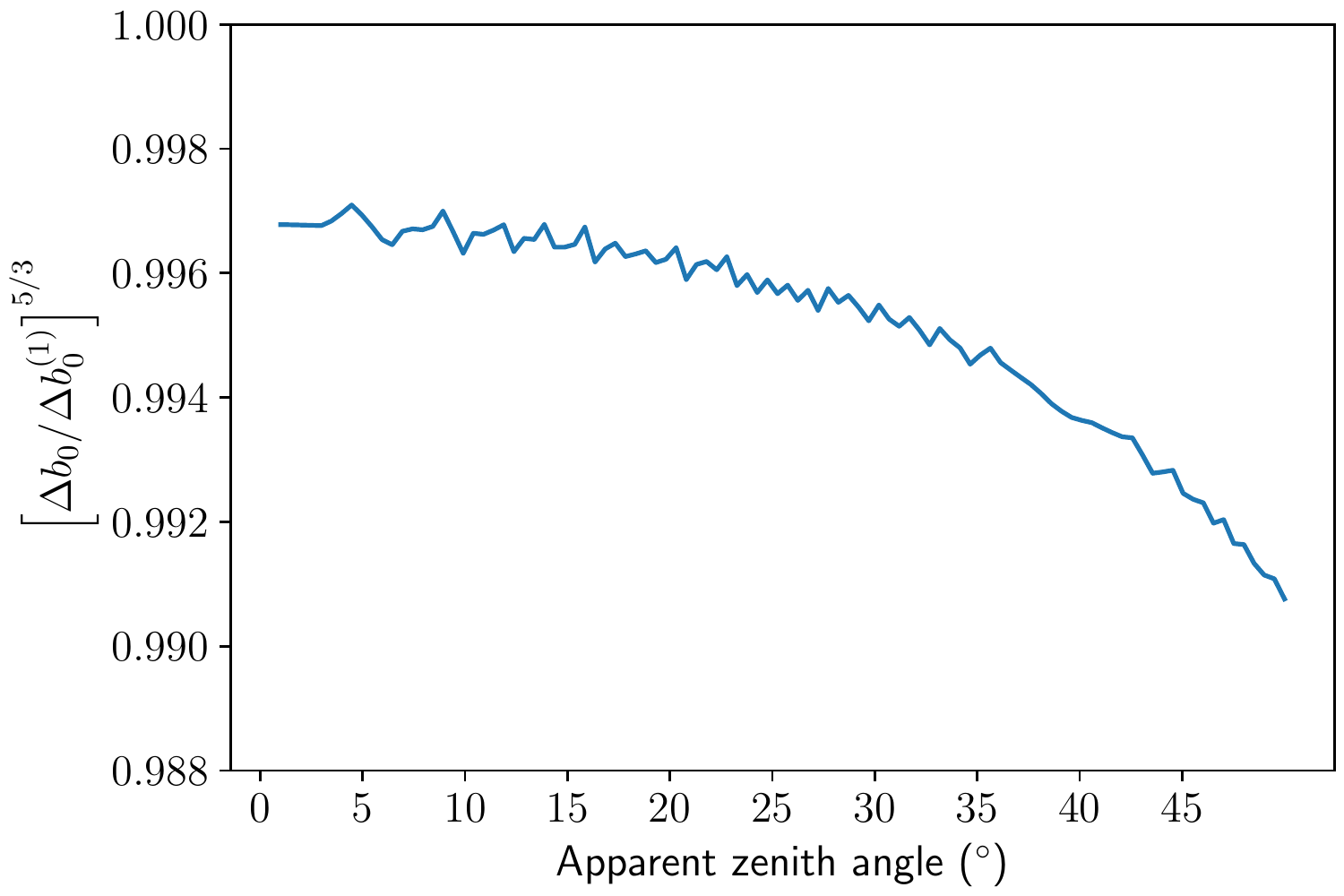}
    \caption{Ratio $\Delta b_0 / \Delta b_0 ^{(1)}$ as a function of zenith angle.}
    \label{fig:b_AO_rapport}
\end{figure}

Next, we take a closer look at the effect of numerical integration. According to Eq.~(10) of \cite{nakajima_zenith-distance_2006}, the phase variance of the wavefront is proportional to $(\Delta b_0)^{5/3}$. In Fig.~   \ref{fig:b_AO_rapport}, we plot the ratio between $\Delta b_0$, calculated using the estimator $b^{(\infty)}$, and $\Delta b_0 ^{(1)}$, calculated using the first-order $b^{(1)}$, all to the power $5/3$ and for $\lambda_\ttt{SCI} = 1630\,\ttt{nm}$.  
In Fig.~\ref{fig:b_AO_rapport}, we notice that the first-order approximation overestimates the values of the lateral displacement of the rays as the ratio is less than 1 regardless of the zenith angle. Also, the ratio studied decreases when the angle $z_0$ increases, but it is still very close to 1, down to 99.1\% at $z_0 = 50^\circ$. Given that the relative lateral deviation $\Delta b_0$ at $45^\circ$ is roughly $0.1\,\ttt{m}$, the difference between the two estimators is approximately $1\,\ttt{mm}$, which is quite negligible with respect to the common Fried parameters ($10\,\ttt{cm}$). The deviation is certainly higher for larger zenith angles, but those are generally avoided when correcting turbulence effects.

Based on Fig.~\ref{fig:b_AO_rapport}, we can reasonably say that in this case, the first-order approximation is quite sufficient because there is little difference between the phase variance calculated with either the Runge-Kutta estimator or with the first-order one. Therefore, the analyses made in the literature (by \cite{wallner_effects_1976, sasiela_strehl_1992, nakajima_zenith-distance_2006}) regarding the impact of the lateral shift on adaptive optics are completely consistent, since we validate the expression of $b^{(1)}$ as a good approximation, even in the more realistic case of a spherically layered atmosphere.

\subsection{Meteor trajectography}\label{sec:lateral_shift_nearbyObjects}

Sometimes one may want to observe objects less than $100 ~\ttt{km}$ away from Earth. It is the case in meteor observations with the recent creation of national and international networks of cameras dedicated to photographing meteorite falls such as the Fireball Recovery and Inter Planetary Observation Network (FRIPON \citep{colas_fripon_2014,colas_fripon_2020}. Usually, the goal of those projects is to estimate the landing position of the object from pictures taken at different locations by several cameras.
\begin{figure}[!ht]
    \centering
    \includegraphics[width=0.45\textwidth,trim={5cm 0cm 35cm 0cm},clip]{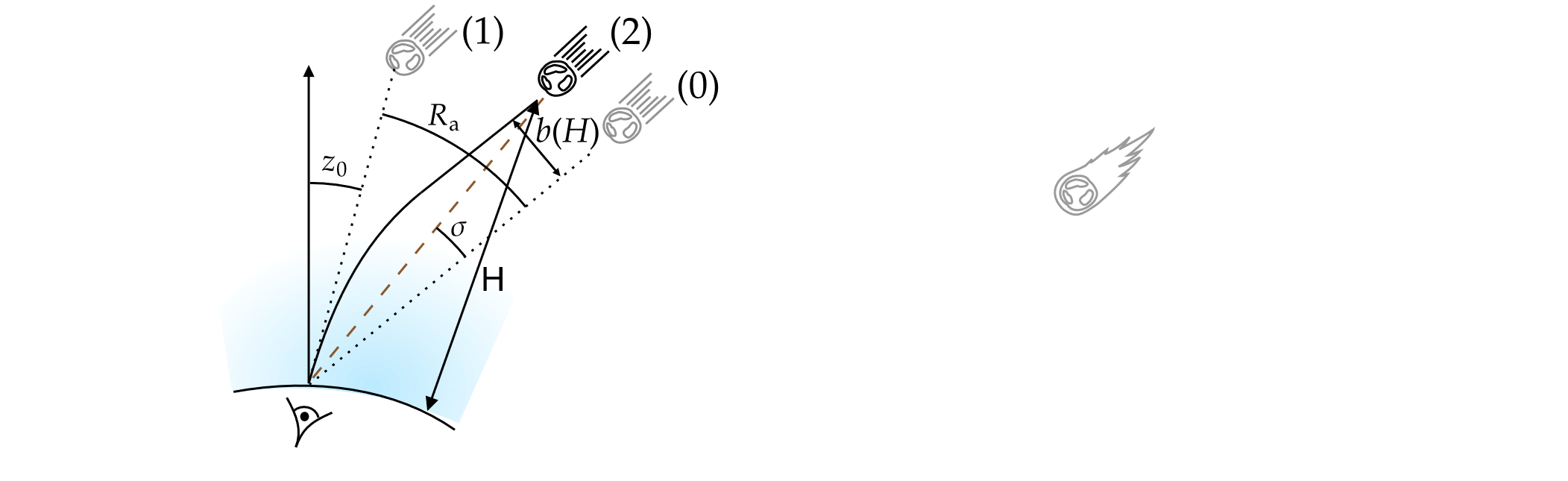}
    \caption{{Difference between the apparent (1) and true (2) positions of a meteor entering Earth's atmosphere.}}
    \label{fig:refraction_4-2}
\end{figure}

As mentioned in the introduction, the all-sky cameras used in meteor tracking networks rely on the astronomical catalogues to calibrate the position of objects in the sky. Thereby, the position of the object is corrected by an angle equal to the refraction angle $R_\ttt{a}$ (Fig.~\ref{fig:refraction_4-2}), whereas the angular correction to be applied would in fact be $R_a - \sigma$, where the angle $\sigma$ is shown in Fig.~\ref{fig:refraction_4-2}; this `refractive parallax' was previously highlighted by \cite{mccrosky_special_1968}. To illustrate, we take a look at Fig.~\ref{fig:refraction_4-2}, where the path of a light ray emitted by a meteorite when it enters Earth's atmosphere is
drawn. The apparent position of the meteorite corresponds to the apparent position of the faraway star (tag $(1)$). When this direction is corrected using the refraction angle $R_\mrm{a}$, we place the object virtually at the tag $(0)$ and make an error equal to the shift $b$.

To quantify the difference between these two corrections and evaluate the error of using only the refraction angle, we can write the compensation by the lateral shift as a correction through the angle $\sigma$ defined in Fig.~\ref{fig:refraction_4-2} such as 
\begin{equation}\label{equ:sigma_1}
    \sin{\sigma} = \frac{b(H)}{l(H)},
\end{equation}
where $l(H)$ is the true distance between the object and the observer and $H$ is the altitude of the object, which is assumed to be available along with the apparent zenith angle $z_0$. 

\begin{figure}[!ht]
    \centering
    \includegraphics[width=0.49\textwidth,trim={0cm 0cm 0cm 0cm},clip]{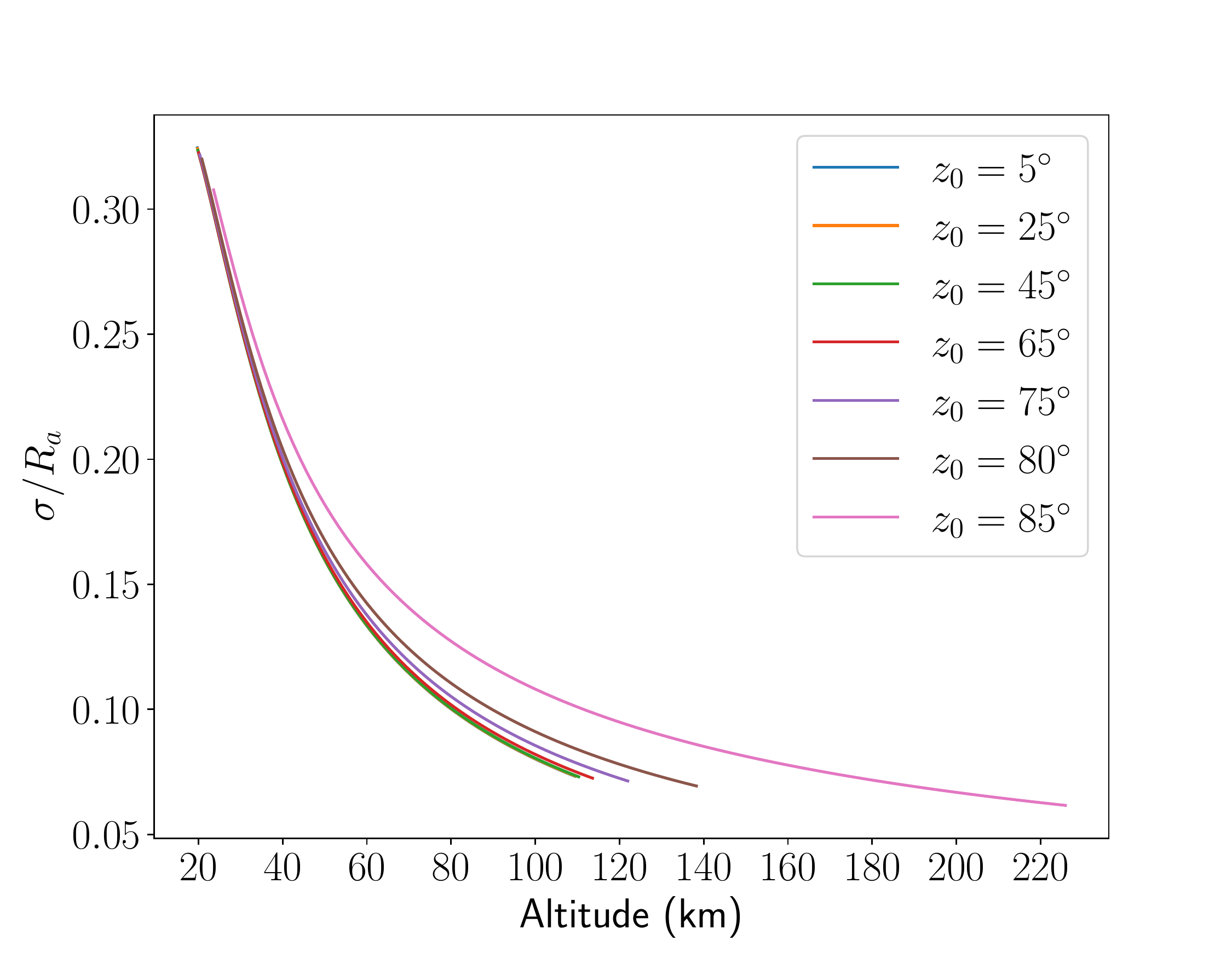}
    \caption{Correction ratio $\sigma/R_\ttt{a}$ versus object's altitude ($H$).}
    \label{fig:sigma_Ra}
\end{figure}

In order to know whether this additional correction is significant or not in relation to the main refraction angle, in Fig.~\ref{fig:sigma_Ra} we plot the evolution of the ratio $\sigma/R_\ttt{a}$ according to altitude. \mLE{The} distance $l$ as a function of the object's altitude $H$, Earth's radius $R_\ttt{T,}$ and the angle $\theta$ defined in Sect.~\ref{sec:numerical_method} \mLE{is given by}:

\begin{equation}\label{equ:distance_D_fripon}
    l^2 = R_\ttt{T} ^2 + (R_\ttt{T} + H)^2 - 2 R_\ttt{T} \left( R_\ttt{T} + H \right) \cos{\theta} .
\end{equation}
In order to compute the values of $\theta$ and $b$ for an object located inside the atmosphere (at an altitude below $H_\ttt{max}$), the Runge-Kutta integration exposed in Sect.~\ref{sec:numerical_method} should be made between the observer level and the altitude of the object (and no longer the atmosphere limit $H_\ttt{max}$). The angle $\theta$ is an output of the resolution of the first set of equations:~(\ref{equ:sys1}) to~(\ref{equ:sys4}).

Based on the reported observations of meteors \citep{jeanne_calibration_2019,gardiol_cavezzo_2021, carbognani_case_2020}, $80~\ttt{km}$ is the typical altitude at which a meteor begins to be observed, and it disappears around an altitude of $20~\ttt{km}$ as it comes apart in the atmosphere. The observation site is assumed at sea level, under the SCTP and with an observation wavelength $\lambda_0$.

We notice in Fig.~\ref{fig:sigma_Ra} that the closer the object gets to Earth, the more significant the correction of angle $\sigma$ becomes; it even reaches $1/3$ of the refraction angle $R_\ttt{a}$ at $H = 20~\ttt{km}$. Also, we can see that the ratio depends only slightly on the zenith angle for small-to-moderate angles and then increases strongly for large zenith angles ($z_0 = 85^\circ$) at a given altitude. As an example of $\sigma$ values, the angular error on the position of an object at an altitude of $40\,\ttt{km}$ and a zenith angle of $80^\circ$ (most meteors are observed at very low elevations) is equal to ${1.04}\, \ttt{arcminutes}$ ($0.2\,R_a$ and $R_a(z_0 = 80^\circ) = {5.2 \,\ttt{arcminutes}}$). {Moreover, if we are interested in the length deviation on the position of the object in the sky, it is simply the lateral shift induced by the bending of the ray from the object to the observer. As the object is no longer at infinity, the value of this deviation is always lower than the lateral shift integrated over the whole atmosphere, previously plotted in Fig.~\ref{fig:sctp_shift}; it is therefore at worst equal to $10\,\mathrm{m}$ at $z_0 = 60^\circ$ and $70 \,\mathrm{m}$ at $z_0 = 80^\circ$. } 

Regarding the specifications of meteor trajectography networks, they usually use all-sky cameras that have a spatial resolution of $10\,\ttt{arcminutes}$ per pixel \citep{gardiol_cavezzo_2021} and do not offer the possibility of taking into account the angular correction $\sigma$. However, many networks have improved this resolution by post-processing. For instance, the CABERNET project \citep{egal_challenge_2017} achieved a precision of $3.24 \, \ttt{arcminutes}$ using photographic records. Also, the fish-eye lenses of FRIPON, \cite{jeanne_calibration_2019} claim to reach an accuracy of $2 \, \ttt{arcminutes}$ for the positions of the stars by improving their distortion models. Compared to these two values, the additional correction $\sigma$ is no less significant and should be taken into account to improve the performance of the detection systems.

A second type of (better resolved) tracking network is able to estimate the position of meteorites with errors lower than $1\,\ttt{arcminute,}$ such as the CAMO network \citep{vida_high_2021}. These networks, which are less dense and certainly more expensive, are dedicated to the calculation of the velocity of meteors, but if scientists plan to use them to estimate the position, then the correction of the lateral shift becomes crucial. 

\section{Summary and conclusion}
In this work, the lateral shift in a spherical Earth approximation with a standard atmosphere model is derived for the first time, to the best of our knowledge. Values of the lateral shift are small for small zenith angles and quickly rise with a zenith angle close to $90^\circ$, as is expected from the known behaviour of the refraction angle. The shift is around $1\,\ttt{m}$ at $20^\circ$, $10\,\ttt{m}$ at $60^\circ,$ and $2000\,\ttt{m}$ at $90^\circ$. The main meteorological factor that impacts the lateral shift was found to be pressure, while the variation as a function of wavelength, which is critical for several applications, does not exceed 2\% of its value at $550\,\ttt{nm}$ over the visible and near-infrared spectrum. The lateral shift also depends on the altitude of the observation site. Its value calculated at an altitude of $4000\,\ttt{m}$ is almost half of its value calculated at sea level.

In order to numerically compute the shift for the largest range of zenith angles, four estimators have been derived. Table~\ref{tab:estimators_conclusion} summarises their accuracy and corresponding equations. A source code implemented using \textsc{Python 3.2} is shared on the collaborative platform GitHub under the name RefractionShift\footnote{See footnote \ref{footnote:github}} and under the open-source General Public License (3.0).

   \begin{table}[!ht]\centering
      \caption[]{The four lateral shift estimators and their zenith range.}
         \label{tab:estimators_conclusion}
     $
         \begin{array}{*{4}{c}}
            \hline \hline
            \noalign{\smallskip}
            \text{Estimator} & \text{Equation}  & \epsilon_\ttt{r}^{(X)} \le 1\% & \epsilon_\ttt{r}^{(X)} \le 0.1\% \\
            \noalign{\smallskip}
            \hline
            \noalign{\smallskip}
            b^{(\infty)} \, \, (\equiv b_0)             & (\ref{equ:system_refraction_ds_2}) &\text{---} &\text{---}\\
            b^{(1)}  & (\ref{equ:b_wallner})\,\&\,  (\ref{equ:approx_shift_1}) & 0-55^\circ & < 1^\circ   \\ 
            b^{(3/2)}     & (\ref{equ:b32}) & 0-70^\circ  & 0-55^\circ  \\ 
            b^{(2)}    & (\ref{equ:b2_approx}) & 0-75^\circ & 0-70^\circ \\  
            \noalign{\smallskip}
            \hline
         \end{array}
     $
   \end{table}

For small-to-moderate zenith angles ($z_0$ smaller than $55^\circ$), the simple closed-form estimator $b^{(1)}$ (previously found by \cite{wallner_effects_1976}) can be used. This a posteriori validates the analyses and computations made for wavefront sensing based on the first-order approximation, in the regime where they are most likely to occur.
For large zenith angles (down to horizon), the Runge-Kutta estimator $b^{(\infty)}$ offers the highest accuracy and is the one to be used for short distance astrometry, such as high-precision meteor tracking. 

Lastly, for zenith angles up to $75^\circ$, a closed-form estimator $b^{(2)}$ is introduced. Analogously to the well-known Laplace formula for the angle of refraction, this general-purpose estimator can further be approximated in most cases by the estimator $b^{(3/2)}$, which enables a simple evaluation of the chromatic behaviour of the shift.


\begin{acknowledgements}
The work of all authors has been funded by \textit{Direction Scientifique Générale} of ONERA in the framework of the \textsc{Costello} project.
The PhD work of H. Labriji is co-funded by Paris-Saclay University. {We thank the reviewer for constructive criticism and valuable comments that helped to improve this paper. We also thank the language editor, N. Saint-Geniès, for corrections that helped improve this paper.}

\end{acknowledgements}

%
\bibliographystyle{aa} 
\bibliography{refraction_article} 

\begin{thebibliography}{58}
\expandafter\ifx\csname natexlab\endcsname\relax\def\natexlab#1{#1}\fi

\bibitem[{Abshire \& Gardner(1985)}]{abshire_atmospheric_1985}
Abshire, J.~B. \& Gardner, C.~S. 1985, IEEE Transactions on Geoscience and
  Remote Sensing, GE-23, 414, conference Name: IEEE Transactions on Geoscience
  and Remote Sensing

\bibitem[{Anderson {et~al.}(1986)Anderson, Clough, Kneizys, Chetwynd, \&
  Shettle}]{Anderson1986}
Anderson, G., Clough, S., Kneizys, F., Chetwynd, J., \& Shettle, E. 1986, {AFGL
  atmospheric constituent profiles (0-120km)} (Environmental Research Paper No.
  964, AFGL-TR-86-0110 (Air Force Geophysics Lab, Hanscom AFB, Massachusetts,
  1986))

\bibitem[{{Aristotle}(1908)}]{aristotle_works_1908}
{Aristotle}. 1908, Works. {Translated} into {English} under the editorship of
  {W}.{D}. {Ross} (Oxford Clarendon Press)

\bibitem[{Auer \& Standish(2000)}]{auer_astronomical_2000}
Auer, L.~H. \& Standish, E.~M. 2000, The Astronomical Journal, 119, 2472

\bibitem[{Barrell {et~al.}(1939)Barrell, Sears, \&
  Bragg}]{barrell_refraction_1939}
Barrell, H., Sears, J.~E., J., \& Bragg, W.~L. 1939, Philosophical Transactions
  of the Royal Society of London. Series A, Mathematical and Physical Sciences,
  238, 1

\bibitem[{Borovicka {et~al.}(1995)Borovicka, Spurny, \&
  Keclikova}]{borovicka_new_1995}
Borovicka, J., Spurny, P., \& Keclikova, J. 1995, Astronomy and Astrophysics
  Supplement Series, 112, 173

\bibitem[{Carbognani {et~al.}(2020)Carbognani, Barghini, Gardiol, Martino,
  Valsecchi, Trivero, Buzzoni, Rasetti, Selvestrel, Knapic, Londero, Zorba,
  Volpicelli, Carlo, Vaubaillon, Marmo, Colas, Valeri, Zanotti, Morini,
  Demaria, Zanda, Bouley, Vernazza, Gattacceca, Rault, Maquet, \&
  Birlan}]{carbognani_case_2020}
Carbognani, A., Barghini, D., Gardiol, D., {et~al.} 2020, The European Physical
  Journal Plus, 135, 255

\bibitem[{Chambers(2005)}]{chambers_astrometry_2005}
Chambers, K. 2005, in Astrometry in the Age of the Next Generation of Large
  Telescopes, Vol. 338, 134

\bibitem[{Ciddor(1996)}]{ciddor_refractive_1996}
Ciddor, P.~E. 1996, Applied Optics, 35, 1566

\bibitem[{COESA(1976)}]{coesa_us_1976}
COESA. 1976, U.{S}. {Standard} {Atmosphere}, 1976, Technical {Memorandum}
  19770009539, NASA Technical Reports Server (NTRS)

\bibitem[{Colas {et~al.}(2020)Colas, Zanda, Bouley, Jeanne, Malgoyre, Birlan,
  Blanpain, Gattacceca, Jorda, Lecubin, Marmo, Rault, Vaubaillon, Vernazza,
  Yohia, Gardiol, Nedelcu, Poppe, Rowe, Forcier, Koschny, Trigo-Rodriguez,
  Lamy, Behrend, Ferrière, Barghini, Buzzoni, Carbognani, Carlo, Martino,
  Knapic, Londero, Pratesi, Rasetti, Riva, Stirpe, Valsecchi, Volpicelli,
  Zorba, Coward, Drolshagen, Drolshagen, Hernandez, Jehin, Jobin, King,
  Nitschelm, Ott, Sanchez-Lavega, Toni, Abraham, Affaticati, Albani, Andreis,
  Andrieu, Anghel, Antaluca, Antier, Appéré, Armand, Ascione, Audureau,
  Auxepaules, Avoscan, Aissa, Bacci, Bǎdescu, Baldini, Baldo, Balestrero,
  Baratoux, Barbotin, Bardy, Basso, Bautista, Bayle, Beck, Bellitto, Belluso,
  Benna, Benammi, Beneteau, Benkhaldoun, Bergamini, Bernardi, Bertaina, Bessin,
  Betti, Bettonvil, Bihel, Birnbaum, Blagoi, Blouri, Boacă, Boatǎ, Bobiet,
  Bonino, Boros, Bouchet, Borgeot, Bouchez, Boust, Boudon, Bouman, Bourget,
  Brandenburg, Bramond, Braun, Bussi, Cacault, Caillier, Calegaro, Camargo,
  Caminade, Campana, Campbell-Burns, Canal-Domingo, Carell, Carreau, Cascone,
  Cattaneo, Cauhape, Cavier, Celestin, Cellino, Champenois, Aoudjehane,
  Chevrier, Cholvy, Chomier, Christou, Cricchio, Coadou, Cocaign, Cochard,
  Cointin, Colombi, Saavedra, Corp, Costa, Costard, Cottier, Cournoyer,
  Coustal, Cremonese, Cristea, Cuzon, D’Agostino, Daiffallah, Dǎnescu,
  Dardon, Dasse, Davadan, Debs, Defaix, Deleflie, D’Elia, Luca, Maria,
  Deverchère, Devillepoix, Dias, Dato, Luca, Dominici, Drouard, Dumont,
  Dupouy, Duvignac, Egal, Erasmus, Esseiva, Ebel, Eisengarten, Federici, Feral,
  Ferrant, Ferreol, Finitzer, Foucault, Francois, Frîncu, Froger, Gaborit,
  Gagliarducci, Galard, Gardavot, Garmier, Garnung, Gautier, Gendre, Gerard,
  Gerardi, Godet, Grandchamps, Grouiez, Groult, Guidetti, Giuli, Hello, Henry,
  Herbreteau, Herpin, Hewins, Hillairet, Horak, Hueso, Huet, Huet, Hyaumé,
  Interrante, Isselin, Jeangeorges, Janeux, Jeanneret, Jobse, Jouin, Jouvard,
  Joy, Julien, Kacerek, Kaire, Kempf, Koschny, Krier, Kwon, Lacassagne, Lachat,
  Lagain, Laisné, Lanchares, Laskar, Lazzarin, Leblanc, Lebreton, Lecomte,
  Dû, Lelong, Lera, Leoni, Le-Pichon, Le-Poupon, Leroy, Leto, Levansuu, Lewin,
  Lienard, Licchelli, Locatelli, Loehle, Loizeau, Luciani, Maignan, Manca,
  Mancuso, Mandon, Mangold, Mannucci, Maquet, Marant, Marchal, Marin,
  Martin-Brisset, Martin, Mathieu, Maury, Mespoulet, Meyer, Meyer, Meza,
  Cecchi, Moiroud, Millan, Montesarchio, Misiano, Molinari, Molau, Monari,
  Monflier, Monkos, Montemaggi, Monti, Moreau, Morin, Mourgues, Mousis,
  Nablanc, Nastasi, Niacşu, Notez, Ory, Pace, Paganelli, Pagola, Pajuelo,
  Palacián, Pallier, Paraschiv, Pardini, Pavone, Pavy, Payen, Pegoraro,
  Peña-Asensio, Perez, Pérez-Hoyos, Perlerin, Peyrot, Peth, Pic, Pietronave,
  Pilger, Piquel, Pisanu, Poppe, Portois, Prezeau, Pugno, Quantin, Quitté,
  Rambaux, Ravier, Repetti, Ribas, Richard, Richard, Rigoni, Rivet, Rizzi,
  Rochain, Rojas, Romeo, Rotaru, Rotger, Rougier, Rousselot, Rousset, Rousseu,
  Rubiera, Rudawska, Rudelle, Ruguet, Russo, Sales, Sauzereau, Salvati,
  Schieffer, Schreiner, Scribano, Selvestrel, Serra, Shengold, Shuttleworth,
  Smareglia, Sohy, Soldi, Stanga, Steinhausser, Strafella, Mbaye, Smedley,
  Tagger, Tanga, Taricco, Teng, Tercu, Thizy, Thomas, Tombelli, Trangosi,
  Tregon, Trivero, Tukkers, Turcu, Umbriaco, Unda-Sanzana, Vairetti,
  Valenzuela, Valente, Varennes, Vauclair, Vergne, Verlinden, Vidal-Alaiz,
  Vieira-Martins, Viel, Vîntdevarǎ, Vinogradoff, Volpini, Wendling, Wilhelm,
  Wohlgemuth, Yanguas, Zagarella, \& Zollo}]{colas_fripon_2020}
Colas, F., Zanda, B., Bouley, S., {et~al.} 2020, Astron. Astrophys., 644, A53

\bibitem[{Colas {et~al.}(2014)Colas, Zanda, Bouley, Vaubaillon, Vernazza,
  Gattacceca, Marmo, Audureau, Kwon, Maquet, Rault, Birlan, Egal, Rotaru,
  Birnbaum, Cochard, \& Thizy}]{colas_fripon_2014}
Colas, F., Zanda, B., Bouley, S., {et~al.} 2014, in Proceedings of the
  {International} {Meteor} {Conference}, {Giron}, {France}, 18-21 {September}
  2014, 34--38

\bibitem[{Devaney {et~al.}(2008)Devaney, Goncharov, \&
  Dainty}]{devaney_chromatic_2008}
Devaney, N., Goncharov, A.~V., \& Dainty, J.~C. 2008, Applied Optics, 47, 1072

\bibitem[{Dodson(1986)}]{dodson_refraction_1986}
Dodson, A.~H. 1986, International Journal of Remote Sensing, 7, 515

\bibitem[{Edlén(1953)}]{edlen_dispersion_1953}
Edlén, B. 1953, JOSA, 43, 339

\bibitem[{Edlén(1966)}]{edlen_refractive_1966}
Edlén, B. 1966, Metrologia, 2, 71

\bibitem[{Egal {et~al.}(2017)Egal, Gural, Vaubaillon, Colas, \&
  Thuillot}]{egal_challenge_2017}
Egal, A., Gural, P.~S., Vaubaillon, J., Colas, F., \& Thuillot, W. 2017,
  Icarus, 294, 43

\bibitem[{Gardiol {et~al.}(2021)Gardiol, Barghini, Buzzoni, Carbognani,
  Di Carlo, Di Martino, Knapic, Londero, Pratesi, Rasetti, Riva, Salerno,
  Stirpe, Valsecchi, Volpicelli, Zorba, Colas, Zanda, Bouley, Jeanne, Malgoyre,
  Birlan, Blanpain, Gattacceca, Lecubin, Marmo, Rault, Vaubaillon, Vernazza,
  Affaticati, Albani, Andreis, Ascione, Avoscan, Bacci, Baldini, Balestrero,
  Basso, Bellitto, Belluso, Benna, Bernardi, Bertaina, Betti, Bonino, Boros,
  Bussi, Carli, Carriero, Cascone, Cattaneo, Cellino, Colombetti, Colombi,
  Costa, Cremonese, Cricchio, D’Agostino, D’Elia, De Maio, Demaria,
  Di Dato, Di Luca, Federici, Gagliarducci, Gerardi, Giuli, Guidetti,
  Interrante, Lazzarin, Lera, Leto, Licchelli, Lippolis, Manca, Mancuso,
  Mannucci, Masi, Masiero, Meucci, Misiano, Moggi Cecchi, Molinari, Monari,
  Montemaggi, Montesarchio, Monti, Morini, Nastasi, Pace, Pardini, Pavone,
  Pegoraro, Pietronave, Pisanu, Pugno, Repetti, Rigoni, Rizzi, Romeni, Romeo,
  Rubinetti, Russo, Salvati, Selvestrel, Serra, Simoncelli, Smareglia, Soldi,
  Stanga, Strafella, Suvieri, Taricco, Tigani Sava, Tombelli, Trivero,
  Umbriaco, Vairetti, Valente, Volpini, Zagarella, \&
  Zollo}]{gardiol_cavezzo_2021}
Gardiol, D., Barghini, D., Buzzoni, A., {et~al.} 2021, Mon. Not. R. Astron.
  Soc., 501, 1215

\bibitem[{Gardner(1977)}]{gardner_correction_1977}
Gardner, C.~S. 1977, Applied Optics, 16, 2427

\bibitem[{Gladstone \& Dale(1863)}]{gladstone_xiv_1863}
Gladstone, J.~H. \& Dale, T.~P. 1863, Philosophical Transactions of the Royal
  Society of London, 153, 317

\bibitem[{Hohenkerk \& Sinclair(1985)}]{hohenkerk_computation_1985}
Hohenkerk, C.~Y. \& Sinclair, A.~T. 1985, The {Computation} of an {Angular}
  {Atmospheric} {Refraction} at {Large} {Zenith} {Angles}, Tech. Rep.~63, HM
  Nautical Almanac Office, Taunton

\bibitem[{Jeanne {et~al.}(2019)Jeanne, Colas, Zanda, Birlan, Vaubaillon,
  Bouley, Vernazza, Jorda, Gattacceca, Rault, Carbognani, Gardiol, Lamy,
  Baratoux, Blanpain, Malgoyre, Lecubin, Marmo, \&
  Hewins}]{jeanne_calibration_2019}
Jeanne, S., Colas, F., Zanda, B., {et~al.} 2019, Astron. Astrophys., 627, A78

\bibitem[{Jolissaint \& Kendrew(2010)}]{jolissaint_modeling_2010}
Jolissaint, L. \& Kendrew, S. 2010, in 1st {AO4ELT} conference - {Adaptive}
  {Optics} for {Extremely} {Large} {Telescopes}, Paris, France, 05021

\bibitem[{Kovalevsky \& Seidelmann(2004)}]{kovalevsky_fundamentals_2004}
Kovalevsky, J. \& Seidelmann, P.~K. 2004, Fundamentals of {Astrometry}
  (Cambridge: Cambridge University Press)

\bibitem[{Kristensen(1998)}]{kristensen_astronomical_1998}
Kristensen, L.~K. 1998, Astronomische Nachrichten, 319, 193

\bibitem[{Li {et~al.}(2016)Li, Zhang, Tang, \& Huang}]{li_method_2016}
Li, S., Zhang, G., Tang, X., \& Huang, W. 2016, Remote Sensing Letters, 7, 985

\bibitem[{Lovchy(2021)}]{lovchy_calculation_2021}
Lovchy, I.~L. 2021, Journal of Optical Technology, 88, 60

\bibitem[{Mahan(1962)}]{mahan_astronomical_1962}
Mahan, A.~I. 1962, Applied Optics, 1, 497

\bibitem[{Marini(1972)}]{marini_correction_1972}
Marini, J.~W. 1972, Radio Science, 7, 223

\bibitem[{Mathar(2007)}]{mathar_refractive_2007}
Mathar, R.~J. 2007, Journal of Optics A: Pure and Applied Optics, 9, 470

\bibitem[{McCrosky \& Posen(1968)}]{mccrosky_special_1968}
McCrosky, R.~E. \& Posen, A. 1968, SAO Special Report, 273

\bibitem[{Nakajima(2006)}]{nakajima_zenith-distance_2006}
Nakajima, T. 2006, The Astrophysical Journal, 652, 1782

\bibitem[{Nauenberg(2017)}]{nauenberg_atmospheric_2017}
Nauenberg, M. 2017, Publications of the Astronomical Society of the Pacific,
  129, 044503

\bibitem[{Néda \& Volkan(2002)}]{neda_flatness_2002}
Néda, Z. \& Volkan, S. 2002, American Journal of Physics, 71

\bibitem[{Owens(1967)}]{owens_optical_1967}
Owens, J.~C. 1967, Applied Optics, 6, 51

\bibitem[{Owner-Petersen(2006)}]{owner-petersen_effects_2006}
Owner-Petersen, M. 2006, in Advances in {Adaptive} {Optics} {II}, Vol. 6272,
  62722F

\bibitem[{Owner-Petersen \& Goncharov(2004)}]{owner-petersen_consequences_2004}
Owner-Petersen, M. \& Goncharov, A. 2004, Proceedings of SPIE - The
  International Society for Optical Engineering

\bibitem[{Pannetier {et~al.}(2021)Pannetier, Mourard, Cassaing, Lagarde,
  Le~Bouquin, Monnier, Sturmann, \&
  Ten~Brummelaar}]{pannetier-longDispCorrection-mnras21}
Pannetier, C., Mourard, D., Cassaing, F., {et~al.} 2021, Mon. Not. R. Astron.
  Soc., 507, 1369–1380

\bibitem[{Radau(1882)}]{radau_recherches_1882}
Radau, R. 1882, Annales de l'Observatoire de Paris, B.1

\bibitem[{Regiomontanus(1544)}]{regiomontanus_scripta_1544}
Regiomontanus, J. 1544, Scripta de {Torqueto}, {Astrolabio} armillari, {Regula}
  magna {Ptolemaica} ... aucta necessariis, {Joannis} {Schoneri} {Carolostadii}
  additionibus ... {Item}, {Libellus} {M}. {Georgii} {Purbachii} de {Quadrato}
  {Geometrico} (Jo. Montanus)

\bibitem[{Saastamoinen(1972)}]{saastamoinen_contributions_1972}
Saastamoinen, J. 1972, Bulletin Géodésique (1946-1975), 105, 279

\bibitem[{Sasiela(1992)}]{sasiela_strehl_1992}
Sasiela, R.~J. 1992, Journal of the Optical Society of America, 9, 1398

\bibitem[{Schmid(1963)}]{schmid_influence_1963-1}
Schmid, H. 1963, The {Influence} of {Atmospheric} {Refraction} on {Directions}
  {Measured} to and from a {Satellite}, {GIMRADA} {Research} note (Army
  Engineer Geodesy, Intelligence and Mapping Research and Development Agency)

\bibitem[{Seidelmann(1992)}]{seidelmann_explanatory_1992}
Seidelmann, P.~K., ed. 1992, Explanatory {Supplement} to the {Astro}- nomical
  {Almanac} (University Science Books), section: 3.28 Refraction

\bibitem[{Skemer {et~al.}(2009)Skemer, Hinz, Hoffmann, Close, Kendrew, Mathar,
  Stuik, Greene, Woodward, \& Kelley}]{skemer_direct_2009}
Skemer, A.~J., Hinz, P.~M., Hoffmann, W.~F., {et~al.} 2009, Publications of the
  Astronomical Society of the Pacific, 121, 897

\bibitem[{Stone(1996)}]{stone_accurate_1996}
Stone, R.~C. 1996, Publications of the Astronomical Society of the Pacific,
  108, 1051

\bibitem[{Thomas \& Joseph(1996)}]{thomas1996astronomical}
Thomas, M.~E. \& Joseph, R.~I. 1996, Johns Hopkins apl technical digest, 17,
  279

\bibitem[{Tóth {et~al.}(2015)Tóth, Kornoš, Zigo, Štefan Gajdoš,
  Kalmančok, Világi, Šimon, Vereš, Šilha, Buček, Galád, Rusňák,
  Hrábek, Ďuriš, \& Rudawska}]{TOTH2015102}
Tóth, J., Kornoš, L., Zigo, P., {et~al.} 2015, Planetary and Space Science,
  118, 102

\bibitem[{van den Born \& Jellema(2020)}]{van_denborn_quantification_2020}
van den Born, J.~A. \& Jellema, W. 2020, Mon. Not. R. Astron. Soc., 496, 4266,
  publisher: Oxford Academic

\bibitem[{van~der Werf(2003)}]{van_der_werf_ray_2003}
van~der Werf, S.~Y. 2003, Applied Optics, 42, 354

\bibitem[{van~der Werf(2008)}]{van_der_werf_comment_2008}
van~der Werf, S.~Y. 2008, Applied Optics, 47, 153

\bibitem[{Vida {et~al.}(2021)Vida, Brown, Campbell-Brown, Weryk, Stober, \&
  McCormack}]{vida_high_2021}
Vida, D., Brown, P.~G., Campbell-Brown, M., {et~al.} 2021, Icarus, 354, 114097

\bibitem[{Wallner(1976)}]{wallner_effects_1976}
Wallner, E.~P. 1976, in Imaging {Through} the {Atmosphere}, Vol. 0075, 119--125

\bibitem[{Wallner(1977)}]{wallner_minimizing_1977}
Wallner, E.~P. 1977, JOSA, 67, 407

\bibitem[{Wehbe {et~al.}(2020)Wehbe, Cabral, \& Avila}]{wehbe_-sky_2020}
Wehbe, B., Cabral, A., \& Avila, G. 2020, Mon. Not. R. Astron. Soc., 499, 183,
  arXiv: 2009.01641

\bibitem[{Wehbe. {et~al.}(2021)Wehbe., Cabral, Sbordone, \&
  Avila}]{wehbe_-sky_2021}
Wehbe., B., Cabral, A., Sbordone, L., \& Avila, G. 2021, Mon. Not. R. Astron.
  Soc., 503, 3818, arXiv: 2103.02273

\bibitem[{Wittmann(1997)}]{wittmann_astronomical_1997}
Wittmann, A.~D. 1997, Astronomische Nachrichten, 318, 305

\bibitem[{Yan {et~al.}(2016)Yan, Wang, Ma, Wang, \& Yu}]{yan_correction_2016}
Yan, M., Wang, C., Ma, J., Wang, Z., \& Yu, B. 2016, Photogrammetric
  Engineering \& Remote Sensing, 82, 427

\end{thebibliography}
%
\begin{appendix}\label{app:appendix}
\section{A model for the dry atmosphere} \label{app:dry_atmosphere}
Below, we present the derivation that leads to the expression of the optical index and its gradient, in the case of dry air and using the temperature profile presented in Section \ref{sec:dry_atm}. We begin by rewriting the Gladstone-Dale relation:
\begin{equation}\label{equ:gladstone_dale}
    n-1 = \kappa\rho .
\end{equation}
Under the ideal gas approximation, and by decoupling the dependencies in $\rho$ and $\lambda$ (respectively, density and wavelength), Eq.~(\ref{equ:gladstone_dale}) reads:
\begin{equation}
    n(\lambda)-1 = A_D(\lambda) \, \frac{P}{T},
\end{equation}
where $A_D$ is called the reduced refractivity of dry air and is given by several models, one of the most recent expressions being given by Eq.~(2) of \cite{ciddor_refractive_1996}. That is,
\begin{align}\label{equ:Ad_ciddor}
    A_D(\lambda) = & 10^{-8} \left[ 5792105 \left(238.0185-1/\lambda^2 \right)^{-1} + \right. \notag \\ 
    & \left. 167917\left(57.362-1/\lambda^2 \right)^{-1} \right] \frac{288.15}{1013.25}.
\end{align}
This expression is given for a wavelength in $\mrm{\mu} \text{m}$ and $A_D$ is in $\text{hPa}^{-1}\text{K}$. The previous expression of $A_D$ applies for wavelengths from $300~\mrm{nm}$ to $1690~\mrm{nm}$ and is also valid over the near-infrared spectrum. 
We then write $P/T$ as a function of the altitude $h$ assuming a constant and known temperature gradient in the troposphere $\omega$ and nil temperature gradient above the tropopause. Hence, we have
\begingroup
\begin{equation}\label{equ:temperature_gradient}
  T(h) =  \begin{cases}
 T_0 + \omega \, h \text{ for }0 \leq h \leq H_{t}, \\
 T_0 + \omega \, H_\ttt{t} \text{ otherwise.}
    \end{cases}
\end{equation}
\endgroup
$H_\ttt{t}$ is the altitude of the tropopause. In this framework, the pressure at a given altitude $h \leq H_\ttt{t}$ depends on the temperature as follows:
\begin{equation}
P(h) = P_0~\left( \frac{T(h)}{T_0}\right)^\gamma,
\end{equation}where 
\begin{equation}\label{equ:gamma_def}
        \gamma = -\dfrac{g M_{\ttt{D}}}{R\omega},
\end{equation}
and
\begin{itemize}
    \item $M_\ttt{D}$  is the molar mass of dry air,  
    \item $\omega := {\mathrm{d}T}/{\mathrm{d}h} = -6.5 \text{ K/km}$ is the temperature gradient in the troposphere, and
    \item $R$ is the universal gas constant.
\end{itemize}
While in the stratosphere, assuming a constant temperature, one obtains the following:
\begin{equation}
    P(h) = P_0 ~\left( \frac{T_t}{T_0}\right)^\gamma \exp \left( -\frac{g M_{\ttt{D}}}{RT_t} (h-H_{t}) \right),
\end{equation}
where $T_\ttt{t}$ is the temperature at the tropopause. We take $H_\ttt{t} = 11~\text{km,}$ as in \cite{hohenkerk_computation_1985}.

Hence, the refractive index with respect to height $h$ reads 
\begingroup
\begin{align}\label{equ:n_dryAir}
   & n(h,\lambda) - 1 = A_D(\lambda)\,P_0\notag \\
   & \times\begin{cases}
 \dfrac{T(h)^{\gamma-1}}{T_0^\gamma} \text{ for }0 \leq h \leq H_{t}, \\
 \dfrac{T_t^{\gamma-1}}{T_0^\gamma} \exp \left( -\dfrac{g M_\ttt{D}}{RT_t} (h-H_{t}) \right) \text{ for } h \geq H_\ttt{t},
    \end{cases}
\end{align}
\endgroup
and its derivative is given by
\begingroup
\begin{align}\label{equ:dndh_dryAir}
   & \frac{\mathrm{d}n(h,\lambda)}{\mathrm{d}h} = A_D(\lambda)\,P_0  \notag \\ & \times \begin{cases}
 \omega (\gamma -1) \dfrac{T(h)^{\gamma-2}}{T_0^\gamma}  \text{ for }0 \leq h \leq H_{t}, \\
 -\dfrac{g M_\ttt{D}}{RT_t}  ~ \dfrac{T_t^{\gamma-1}}{T_0^\gamma} \exp \left( -\dfrac{g M_\ttt{D}}{RT_t} (h-H_{t}) \right) \text{for } h\geq H_\ttt{t}.
    \end{cases}
\end{align}
\endgroup

\begin{figure}[!ht]
    \centering
    \includegraphics[width=0.49\textwidth]{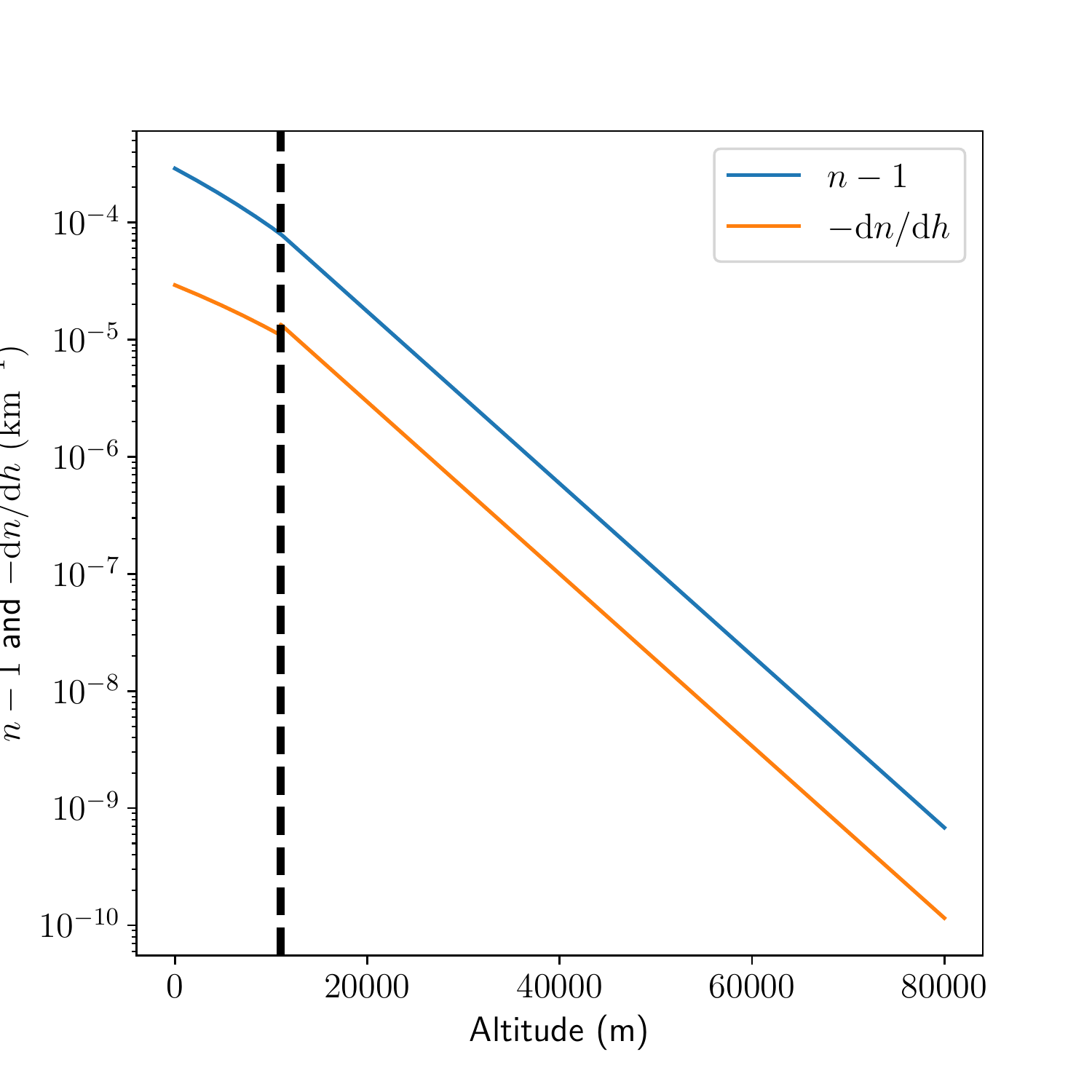}
    \caption{Evolution of refractive index and its derivative according to altitude.{ The dashed line represents the altitude of the tropopause.} }
    \label{fig:model_atm_n_dn}
\end{figure}
Figure~\ref{fig:model_atm_n_dn} shows the evolution of the refractive index and its derivative as a function of altitude. We notice that $n$ and $\mrm{d}n/\mrm{d}h$ decrease exponentially in the stratosphere (the scale of the graph on the $x$ axis is linear and on the $y$ axis logarithmic). Also, we observe a discontinuity of $\mrm{d}n/\mrm{d}h$ at the boundary between the troposphere and the stratosphere, which is due to the discontinuity of the temperature gradient.

Equations~(\ref{equ:n_dryAir})~and~(\ref{equ:dndh_dryAir}) provide us with the evolution of the refractive index $n$ and its derivative $\mrm{d}n/\mrm{d}h$ in the atmosphere, under the chosen temperature and pressure model. In Sect.~\ref{sec:numerical_method}, these two pieces of information allow us to solve the equation system~(\ref{equ:system_refraction_ds_2}) and build a first numerical estimator of the lateral shift.

\section{The main atmospheric moments} \label{app:moment_atmosphere}

 In this paragraph, we define the main characteristic moments of the atmosphere and write them as functions of temperature and pressure using the dry model of Earth's atmosphere. We recall the definitions of the three moments:
\begin{subequations}
    \begin{align}\label{equ:moments_atm}
        L^1(h) & = \int_h^\infty \Bar{\rho} \, \mrm{d}x,\\
        L^2(h) & = \int_h^\infty \Bar{\rho}^2 \, \mrm{d}x,\\
        L^\ttt{b}(h)   & = \left[ \int_h^\infty x\,\Bar{\rho} \, \mrm{d}x \right] / L^1(h).
    \end{align}
\end{subequations}
$L^\ttt{b}(h)$ can be interpreted as the altitude of the barycenter of an air column.  

Using the hydrostatic equilibrium equation, $L^1$ can directly be written as
\begin{equation}\label{equ:L1}
    L^1(h) = \frac{P(h)}{g \rho_0}.
\end{equation}

Besides this, using the atmosphere model exposed in Appendix~\ref{app:dry_atmosphere} we can explicitly express the moments $L^2_0$ and $L^\ttt{b}_0$ as functions of $T_0$, $H_t$, and the ratio of $T_\ttt{t}$ and $T_0$:
\begin{equation}
r := \frac{T_\ttt{t}}{T_0}.
\end{equation}

We obtain the following:
\begin{align}
    L^2_0 =& \, \frac{\gamma L^1_0}{2\gamma - 1} - \frac{ L^1_0}{2(2\gamma - 1)} r^{2\gamma - 1} \text{ and} \label{equ:L2}\\
    L^\ttt{b}_0 = & \, h_0  \left(1 - \frac{r^\gamma}{\gamma +1 } \right) + \frac{\gamma }{\gamma +1} L^1_0 \left( 1 - r^\gamma  + \frac{\gamma +1}{\gamma} r^{\gamma +1} \right) \notag \\
    & +  \frac{r^\gamma H_t}{\gamma +1} \label{equ:Lb}.
\end{align}

The evolution of $L^1$ $L^2$ and $L^\mrm{b}$ as a function of the altitude is shown in Fig. \ref{fig:longueur_approx2}. In order to obtain this figure, we changed both the altitude and the temperature and pressure values at the observation site. The evolution of $P(h_0)$ and $T(h_0)$ is based on the model of the dry atmosphere exposed in Appendix~ \ref{app:dry_atmosphere} using the standard values of $P(h_0 = 0)$ and $T(h_0 = 0)$ (Eq.~(\ref{equ:SCTP})).
Using these compact expressions of the atmospheric moments (Eqs.~(\ref{equ:L1}),~(\ref{equ:L2})~and~(\ref{equ:Lb})), we explicitly express three new estimators of the lateral shift in Sect.~\ref{sec:taylor_expansion}.

\end{appendix}

\end{document}